\documentclass[journal]{IEEEtran}
\IEEEoverridecommandlockouts

\usepackage{cite}
\usepackage{amsmath,amssymb,amsfonts}
\usepackage{algorithmic}
\usepackage{graphicx}
\usepackage{textcomp}
\usepackage{xcolor}
\usepackage{booktabs}
\usepackage{color,soul}
\usepackage{tabularx,hhline}
\usepackage{multirow}
\usepackage{hyperref}
\usepackage{tikz}
\usepackage{textcomp}
\usepackage{enumitem}

\usepackage[strings]{underscore}
\usepackage[left=1.62cm,right=1.62cm,top=0.7in]{geometry}
\usepackage{algorithm}
\usepackage{eurosym}
\usepackage{subfigure}
\usepackage{pbalance}

\begin{document}

\title{
A Cyber Insurance Policy for Hedging Against Load-Altering Attacks and Extreme Load Variations in Distribution Grids}

\author{Shijie Pan, Zaint A. Alexakis, Subhash~Lakshminarayana, Charalambos Konstantinou

\thanks{{S. Pan, Z. A. Alexakis, and C. Konstantinou are with the Computer, Electrical and Mathematical Sciences and Engineering (CEMSE) Division, King Abdullah University of Science and Technology (KAUST), Thuwal 23955-6900, Saudi Arabia. S. Lakshminarayana is with the School of Engineering, University of Warwick, Coventry, UK, CV47AL. }}

}

\IEEEaftertitletext{\vspace{-2.2\baselineskip}}
\maketitle
\begin{abstract} 

Uncertainties in renewable energy resources (RES) and load variations can lead to elevated system operational costs. Moreover, the emergence of large-scale distributed threats, such as load-altering attacks (LAAs), can induce substantial load variations, further exacerbating these costs. Although traditional defense measures can reduce the likelihood of such attacks, considerable residual risks remain. Thus, this paper proposes a cyber insurance framework designed to hedge against additional operational costs resulting from LAAs and substantial load variations in renewable-rich grids. The insurance framework determines both the insurance coverage and premium based on the Value at Risk (VaR) and Tail Value at Risk (TVaR). These risk metrics are calculated using the system failure probability and the probability density function (PDF) of the system operation cost. The system failure probability is assessed through a semi-Markov process (SMP), while the cost distribution is estimated through a cost minimization model of a distribution grid combined with a Monte Carlo simulation to capture load variability. Furthermore, we employ a bi-level optimization scheme that identifies the specific load distribution leading to the maximum system cost, thereby enhancing the accuracy of the operation cost PDF estimation. The effectiveness and scalability of the proposed cyber insurance policy are evaluated considering a modified IEEE 118-bus test system and the IEEE European low-voltage (LV) test feeder model. The case study shows that with a relatively low premium, the network operator can hedge against additional operational costs caused by malicious load manipulations.
\end{abstract}

\begin{IEEEkeywords}
Distribution grid, load curtailment, optimization, cyber insurance, semi-Markov process.
\end{IEEEkeywords}

\section*{Nomenclature}
\addcontentsline{toc}{section}{Nomenclature}
\begin{IEEEdescription}[\IEEEusemathlabelsep\IEEEsetlabelwidth{$V_1,V_2,V_3$}]

\item[\textit{Sets and Indexes}] 
\item[$\mathcal{T}$] Set of hours in a day, indexed by $t$.
\item[$\mathcal{B}$] Set of buses, indexed by $i$.
\item[$\mathcal{L}$] Set of power lines, indexed by $ij$.

\item[\textit{Variables}] 
\item[$\boldsymbol{C}$] Total operational cost, \euro.
\item[$Q^{PV}_{i,t}$] PV reactive power generation on bus $i$ at time $t$, $MVar$.
\item[$P^{Curtail}_{i,t}$] Curtailed load demand (active power) on bus $i$ at time $t$, $MW$.
\item[$Q^{Curtail}_{i,t}$] Curtailed load demand (reactive power) on bus $i$ at time $t$, $MVar$.
\item[$P^{Load}_{i,t}$] Actual load demand (active power) on bus $i$ at time $t$, $MW$.
\item[$Q^{Load}_{i,t}$] Actual load demand (reactive power) on bus $i$ at time $t$, $MVar$.
\item[$P^{var}_{i,t}$] Load variation (active power) on bus $i$ at time $t$, $MW$.
\item[$Q^{var}_{i,t}$] Load variation (reactive power) on bus $i$ at time $t$, $MVar$.

\item[$P^{Buy}_{i,t}$] Power purchased from the energy market on bus $i$ at time $t$, $MW$.
\item[$P^{Sell}_{i,t}$] Power sold to the energy market on bus $i$ at time $t$, $MW$.
\item[$P^{Market}_{i,t}$] Power flows from the energy market on bus $i$ at time $t$, $MW$.

\item[$P^{Battery}_{i,t}$] Power flows from the battery on bus $i$ at time $t$, $MW$.
\item[$E^{Charge}_{i,t}$] Battery charge on bus $i$ at time $t$, $MW$.
\item[$E^{Discharge}_{i,t}$] Battery discharge on bus $i$ at time $t$, $MW$.

\item[$E^{Remain}_{i,t}$] Remaining energy in the battery on bus $i$ at time $t$, $MW$.

\item[$z_{i,t}$] Binary variable representing BESS charging status on bus $i$ at time $t$.

\item[$p_{ij,t}$] Active power flow on line $ij$ at time $t$, $MW$.
\item[$q_{ij,t}$] Reactive power flow on line $ij$ at time $t$, $MVar$.

\item[$v_{i,t}$] Square of voltage on bus $i$ at time $t$, $V^2$.

\item[\textit{Parameters}] 
\item[$\mathit{\Delta}t$] Time step, $h$.
\item[$\boldsymbol{c}^{EE}$] Electricity expense, \euro.
\item[$\boldsymbol{c}^{LC}$] Load Curtailment Compensation, \euro.
\item[$c^B_t$] Price of buying power from the market, \euro$/MWh$.
\item[$c^S_t$] Price of selling power to the market, \euro$/MWh$.
\item[$P^{Dem}_{i,t}$] Nominal load active power demand on bus $i$ at time $t$, $MW$.
\item[$Q^{Dem}_{i,t}$] Nominal load reactive power demand on bus $i$ at time $t$, $MVar$.
\item[$\hat{P}^{Dem}_{i,t}$]  Adjusted load active power demand by the Monte Carlo simulation, on bus $i$ at time $t$, $MW$.
\item[$\hat{Q}^{Dem}_{i,t}$]  Adjusted load reactive power demand by the Monte Carlo simulation, on bus $i$ at time $t$, $MVar$.
\item[$P^{PV}_{i,t}$] PV active power generation on bus $i$ at time $t$, $MW$.
\item[$S^{PV}_{i,t}$] PV capacity on bus $i$ at time $t$, $MVA$.
\item[$E^{Capacity}_{i,t}$] Battery capacity on bus $i$ at time $t$, $MW$.
\item[$E^{Limit}_{i,t}$] Battery charging/discharging limit on bus $i$ at time $t$, $MW$.
\item[$r_{ij}, x_{ij}$] Resistance and reactance on line $ij$, $\Omega$.
\item[$V_{min},V_{max}$] Upper and Lower limits of bus voltages, $V$.
\item[$\eta_{M}$] Power loss factor of electricity market transactions.
\item[$\eta_{B}$] Battery charging efficiency.
\item[$\gamma$] Maximum load curtailment percentage.
\item[$\delta$] Load uncertainty factor per single bus.
\item[$\Delta$] Load uncertainty factor for the whole system.
\item[$a$] Load curtailment compensation penalty factor.
\item[$b$] One-time incentives for load curtailment, \euro.

\end{IEEEdescription}
\section{Introduction}
\label{section:Introduction}

A noteworthy shift is happening in energy systems globally, with governments worldwide progressively enacting policies to encourage higher penetration of renewable energy sources (RES). For example, Saudi Arabia's NEOM project aims to become the world's first city powered entirely by RES, integrating solar, wind, and green hydrogen technologies~\cite{NEOM}. The power generated by RES, however, is intermittent due to its dependence on weather conditions.
To attenuate this variability, battery energy storage systems (BESS) play a crucial role by accumulating excess energy during periods of surplus and supplying it to the grid during periods of deficit.
Studies have analyzed methods to optimize BESS scheduling in grids with high RES penetrations \cite{liu2024system,bera2023reliability,zhang2022energy}. Additionally, the energy conserved through optimal BESS scheduling can be marketed to other entities, thereby increasing profitability if the grid operator engages in the energy market \cite{lezama2018local,
zhang2024price}.
 
At the same time, strategies such as load curtailment are frequently utilized to decrease demand, serving as an additional method to address the RES's intermittent nature. Incentives are often offered to users as a reward for reducing the demand to encourage their participation. For instance, in \cite{lou2013profit}, the authors propose a profit-optimal and stability-aware load curtailment framework that optimally allocates curtailment levels across buses while minimizing system disruption, offering an integrated perspective on both operational cost and stability under smart grid environments. Elyasichamazkoti \textit{et al.}~\cite{elyasichamazkoti2022optimal} present a mathematical model for under-frequency load curtailment, aiming to minimize the total absolute change in line flows and thereby preserve system operating points within the system security constraints. Similarly, Javed \textit{et al.}~\cite{javed2022quantitative} study the impact of load curtailment on an off-grid RES-based system. Results show that the flexibility introduced by load management can greatly reduce the operational cost by more than 40\%. 

However, the optimization models in these works rely on nominal load profiles, which can be influenced by real-time load variations \cite{CaliEle}. Moreover, as the variety of loads connected to the grid increases, such as electric vehicles (EVs), short-term load changes take place more frequently than before~\cite{gruosso2019probabilistic}. According to a report \cite{wilson2023era}, unexpected surges in electricity demand, driven by new industrial developments, rapid electrification across sectors, and the growing power needs of AI data centers, can challenge existing distribution networks. Electricity demand is forecast to rise from $2.6\%$ to $4.7\%$ over the next five years, with peak demand increasing by $9\%$ in extreme cases.

Although several studies consider the stochastic nature of load profiles~\cite{zhang2024stochastic}, extreme load variations may still occur, resulting in sharp increases in operational costs. Beyond natural stochastic fluctuations, malicious manipulations of loads, known as load-altering attacks (LAAs)~\cite{elyasichamazkoti2022optimal,maleki2025survey}, pose an additional and more severe threat. LAAs constitute a multifaceted risk that spans multiple timescales. In short timescales, they aim to degrade transient performance by disrupting frequency and voltage stability, as discussed in~\cite{maleki2025survey,goodridge2024uncovering}.

Practically, such short-term LAAs can be realized by coordinating high-wattage loads at scale, thereby increasing grid operational stress or even inducing instability. For example, a botnet of compromised high-power devices, such as EV chargers, can be orchestrated to simultaneously alter their consumption and launch large-scale attacks on the grid. These devices often rely on standardized communication protocols, which expand the system’s attack surface, as evidenced by common vulnerabilities and exposures (CVEs) reported in control firmware and communication interfaces. Recent studies~\cite{lakshminarayana2022load,maleki2025survey} have summarized and demonstrated these pathways as realistic channels for executing LAAs.

In a longer time scale, stealthy load manipulations can change the system load distribution and ultimately increase system operating costs. For instance, cyber-attacks targeting heating, ventilation, and air conditioning (HVAC) systems can compromise and manipulate their operation, deliberately increasing or decreasing building power consumption~\cite{chen2023review,chen2025defending}. When a large number of HVAC units are compromised, the distorted demand can shift locational marginal prices, reducing operator profits and even influencing market prices~\cite{barreto2020attacking,ospina2023feasibility}. From the perspective of demand flexibility~\cite{yang2024secure}, although the system can attempt to balance manipulated loads by coordinating other participants, this inevitably leads to suboptimal or infeasible scheduling and further elevates operational costs.

While short-term mitigation and detection strategies have been extensively studied \cite{liu2021robust,lakshminarayana2022data,chu2022mitigating}, addressing the long-term impacts of LAAs on system operation remains largely unexplored. Particularly, risk management is essential for grid operators to mitigate these challenges. Cyber insurance emerges as a solution, which is a contract that entities can purchase to shift part of the financial risks and economic burden associated with cyber-attacks from the insured party to the insurer. In practice, the insured customers pay a premium based on their estimated risk exposure, and in the event of a cyber-attack, the insurer provides compensation up to the agreed coverage limit. In the industry, cyber insurance has been adopted to protect critical infrastructure, with providers offering tailored coverage for operational technology (OT) environments, including energy and utility systems \cite{munichre_cyber_solutions, allstate}. While these policies commonly cover risks such as control system disruptions and ransomware attacks, there is still a noticeable lack of real-world cyber insurance policies specifically designed for hedging financial risks in power system operations. This gap has motivated a growing body of academic research to examine the feasibility and design of cyber insurance mechanisms tailored to the unique challenges of power system operations. Acharya \textit{et al.}~\cite{acharya2021cyber} propose cyber insurance for public EV charging stations to hedge the financial loss due to cyber-attacks via the manipulation of high-wattage appliances. A data-driven cyber insurance design model is formulated to derive an optimal insurance premium. In \cite{lau2022novel}, the authors propose a Shapley mutual cyber insurance principle for substations while also considering the allocation of defense resources.

To date, no studies have explored how a cyber insurance policy can mitigate financial risks associated with load variations. Addressing such events poses a particular challenge due to the dynamic and unpredictable nature of load variability.
In this work, we propose a cyber insurance policy designed to help grid operators hedge against such risks, particularly in the context of high penetration of RES in the distribution grids.
The main contributions of this work are as follows:

\noindent (1) The impact of load variations on the grid's operational cost and load curtailment is studied. We use a Monte Carlo simulation to obtain the operational cost under different load demand uncertainties. A bi-level optimization model is proposed to assess the maximum impact of load variations. The lower level (LL) optimizes BESS charging, energy trading, and load curtailment to minimize the grid's operational cost. The upper level (UL) determines the load variation at each bus, which maximizes the grid's operational cost. The solution of the bi-level problem, combined with the results of the Monte Carlo simulation, is incorporated to accurately estimate the probability density function (PDF) of operational cost. This PDF is then used to determine the insurance coverage. 

\noindent (2) A distribution grid model with $100\%$ RES penetration is considered in which the pricing model takes into account the cost during market participation as well as the cost incurred by load curtailment compensation. We subsequently develop a cost-minimization model from the grid operator's perspective to achieve optimal integration of BESS, facilitate market participation, and manage load curtailment, thereby reducing operational costs. 

\noindent (3) A cyber insurance policy framework is proposed in order to hedge against the potential financial loss caused by extreme load variations and LAAs. A semi-Markov process (SMP) model is used to estimate the probability of these extreme situations. The coverage and premium of the insurance are calculated based on value at risk (VaR) and tail value at risk (TVaR) \cite{lau2021coalitional}. We demonstrate that cyber insurance policies in distribution grids can mitigate the risks associated with high operational costs resulting from load variations with a relatively low premium requirement. 

The rest of the paper is as follows. 
Section~\ref{section:GridModel} presents the formulation of the grid model for which the insurance will be considered. Section~\ref{section:CyberInsurance} studies the uncertainty of load variations and designs the insurance premium and coverage. 
Results are presented in Section~\ref{section:Simulation}, while 
Section~\ref{section:Conclusion} concludes the work.

\section{Grid Model and Load Curtailment}
\label{section:GridModel}

This section introduces the grid and optimization model that serves as the foundation for the proposed cyber insurance framework. The cost function of the grid's operation is first defined, and then, we formulate its cost-minimum optimization problem.

\begin{figure}[t]
\centering    \includegraphics[width=0.38\textwidth]{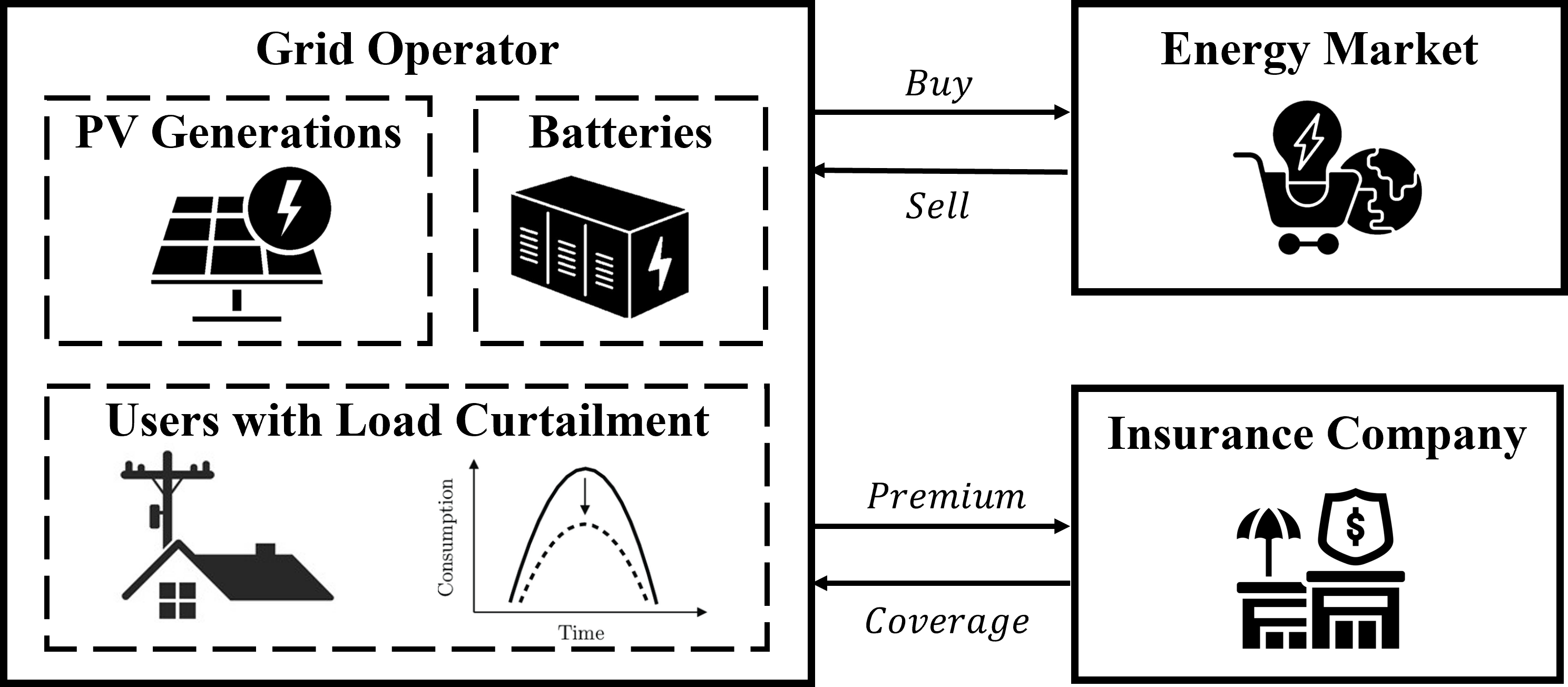}
    \caption{Operational overview of the grid model under consideration.}
    \label{fig:general}
\end{figure}

The objective of the grid operator is to maintain grid operations at a minimum cost. We consider a grid model with 100\% penetration of RES, specifically PV systems, as illustrated in Fig.~\ref{fig:general}. This assumption aligns with national and global energy transition targets, such as Saudi Arabia's Vision 2030 and international decarbonization goals \cite{NEOM}. Moreover, fully renewable-based grids present unique operational and stability challenges, particularly in the absence of conventional synchronous generators, making them a critical research focus in power system dynamics and control. When the electricity supply exceeds demand, the grid operator can sell the surplus electricity to the energy market to generate profit. Conversely, when demand exceeds supply, the operator must either purchase additional power from the market to meet the demand or implement load curtailment. By integrating BESS and engaging with the market, the grid operator can employ smart charging strategies \cite{deb2022smart}. 

Compared to traditional synchronous generators that exhibit a quadratic cost function due to fuel and maintenance expenses, RES incur almost negligible generation costs. Consequently, daily operational costs primarily arise from two sources:
\textbf{Electricity Expense:} 
The grid operator might be required to purchase power from the electricity market during peak hours or sell stored and excess power during other periods for profit. These expenses, referred to as the electricity expense $\boldsymbol{c}^{EE}$, depend on the market price of electricity and the volume of transactions, and can be expressed as: 
\begin{equation} \label{EE}
    \boldsymbol{c}^{EE} = \sum_{t \in \mathcal{T}}\sum_{i \in \mathcal{B}} (c^{B}_{t}P^{Buy}_{i,t}\mathit{\Delta}t-c^{S}_{t}P^{Sell}_{i,t}\mathit{\Delta}t).
\end{equation}
\textbf{Load Curtailment Compensation:} In the scenario where the operator curtails a load, it compensates the user through an incentive mechanism. 
In this work, we consider the incentives as a load curtailment compensation cost $\boldsymbol{c}^{LC}$. It follows the formulation proposed by Lou \textit{et al.}~\cite{lou2013profit}, which is proved to be a convex function of the curtailed load demand $P^{Curtail}$: 
\begin{equation} \label{LC}
    \boldsymbol{c}^{LC} = \sum_{t \in \mathcal{T}} \sum_{i \in \mathcal{B}} a(\sqrt{P^{Dem}_{i,t}}-\sqrt{P^{Dem}_{i,t}-P^{Curtail}_{i,t}}) + b.
\end{equation} 
The parameter $a$ defines the penalty range for curtailment. The constant $b$ represents an one-time incentive offered to users when signing the curtailment agreement, ensuring their profits are not negatively impacted. Specifically, $b$ is set to compensate for the cost saved between the curtailed and non-curtailed solutions, aligning each user's profit with their marginal contribution to operational cost reduction. 

The total operational cost $\boldsymbol{C}$ is the sum of the electricity expense and the load curtailment compensation. An optimization problem is formulated to enable automatic decision-making: 
\begin{equation} \label{Objective}
\min\limits_{\text{$P^{Buy}_{i,t}$, $P^{Sell}_{i,t}$, $P^{Curtail}_{i,t}$,$Q^{PV}_{i,t}$}} \quad \boldsymbol{C} = \boldsymbol{c}^{EE} + \boldsymbol{c}^{LC}
\end{equation} 
Next, we introduce the constraints for each of the participating components in the system. 

\textbf{Energy Market: } We account for proportional power losses during electricity market transactions \cite{paudel2020peer}; thus, the power flows from the energy market can be expressed as:
\begin{equation} \label{Market1}
    P^{Market}_{i,t} = \frac{1}{\eta_{M}}P^{Buy}_{i,t} - \eta_{M}P^{Sell}_{i,t},
\end{equation} 
\begin{equation} \label{Market2}
    P^{Buy}_{i,t},P^{Sell}_{i,t} \geq 0.
\end{equation} 

\textbf{BESS:} Considering the BESS charging efficiency, the power supplied from the BESS to the grid is expressed as:
\begin{equation} \label{Battery1}
    P^{Battery}_{i,t} =  E^{Discharge}_{i,t} - E^{Charge}_{i,t},
\end{equation}
where
\begin{equation} \label{Battery2}
    0 \leq E^{Charge,Discharge}_{i,t} \leq E^{Limit}_{i}.
\end{equation}  
\eqref{BatteryRemain}-\eqref{BatteryLimit} ensure that BESS charging does not exceed its rated capacity, and discharging does not surpass the remaining power:
\begin{equation} \label{BatteryRemain}
    E^{Remain}_{i,t+1} = E^{Remain}_{i,t} - \frac{1}{\eta_{B}}E^{Discharge}_{i,t} + \eta_{B}E^{Charge}_{i,t},
\end{equation}
\begin{equation} \label{BatteryLimit}
    0 \leq E^{Remain}_{i,t} \leq E^{Capacity}_{i}. 
\end{equation}

To prevent the BESS from charging and discharging simultaneously, the following constraint is typically imposed:
\begin{equation} \label{BatteryCD}
    E^{Charge}_{i,t} \cdot E^{Discharge}_{i,t} = 0
\end{equation}This nonlinear constraint ensures that the BESS operates in either charging or discharging mode at any given time. It can be equivalently reformulated into a linear form using a binary variable:
\begin{equation}
   \label{BatteryCD_linear1}
    E^{Charge}_{i,t} \leq E^{Limit}_{i} \, z_{i,t}
\end{equation} 
\begin{equation}
   \label{BatteryCD_linear2}
    E^{Discharge}_{i,t} \leq E^{Limit}_{i} \, (1 - z_{i,t})
\end{equation}Here, \( z_{i,t} \in \{0,1\} \) represents the operational mode: \( z_{i,t} = 1 \) indicates charging, while \( z_{i,t} = 0 \) indicates discharging. This reformulation transforms the original nonlinear condition into a mixed-integer linear constraint that is solver-friendly.

\textbf{Load Curtailment:} With  curtailment, the load demand is: 
\begin{equation}\label{Curtail1} 
    P^{Load}_{i,t} = P^{Dem}_{i,t} - P^{Curtail}_{i,t}.
\end{equation}
\begin{equation}\label{Curtail2} 
    Q^{Load}_{i,t} = Q^{Dem}_{i,t} - Q^{Curtail}_{i,t}.
\end{equation}
To limit load curtailment, we define $\gamma$ as the maximum allowable percentage~\cite{xu2016optimization}: 
\begin{equation} \label{Curtail3}
    0 \leq P^{Curtail}_{i,t} \leq \gamma P^{Dem}_{i,t}.
\end{equation}
\begin{equation} \label{Curtail4}
    -\gamma Q^{Dem}_{i,t} \leq Q^{Curtail}_{i,t} \leq \gamma Q^{Dem}_{i,t}.
\end{equation}

\textbf{PV System:}
The PV systems are considered inverter-based resources that can provide reactive power compensation with negligible costs. Its constraint is expressed as \eqref{PV} to ensure the generated power does not exceed the capacity.
\begin{equation} \label{PV}
     -\sqrt{{S^{PV}_{i,t}}^2-{P^{PV}_{i,t}}^2} \leq Q^{PV}_{i,t} \leq \sqrt{{S^{PV}_{i,t}}^2-{P^{PV}_{i,t}}^2}.
\end{equation}

\textbf{Power Flow Constraints: } For the power flow calculation, we use the LinDistFlow model~\cite{baran2002optimal}, which is a branch flow model for radial distribution networks. \eqref{PowerFlowP}-\eqref{PowerFlowV} are the branch flow and nodal voltage balance equations, and \eqref{VSecurity} limits the system's upper and lower bounds of bus voltages: 
\begin{equation}\label{PowerFlowP}
\begin{aligned}
    p_{ij,t} = &P^{Load}_{j,t}-P^{PV}_{j,t}-P^{Market}_{j,t} -P^{Battery}_{j,t}
    \\ &+ \sum p_{jk,t},
\end{aligned}
\end{equation}
\begin{equation}\label{PowerFlowQ}
\begin{aligned}
    q_{ij,t} = Q^{Load}_{j,t}-Q^{PV}_{j,t} + \sum q_{jk,t},
\end{aligned}
\end{equation}
\begin{equation}\label{PowerFlowV}
\begin{aligned}
    v_{j,t} = v_{i,t}-2(r_{ij}p_{ij,t}+x_{ij}q_{ij,t}).
\end{aligned}
\end{equation}
\begin{equation}\label{VSecurity}
    V_{min}^2 \leq v_{i,t} \leq V_{max}^2,
\end{equation}

Incorporating the above constraints, the optimization model of the grid can be formulated as shown below:
\begin{equation} 
\begin{aligned}
\min\limits_{\text{$P^{Buy}_t$, $P^{Sell}_t$, $P^{Curtail}_{i,t}$,$Q^{PV}_{i,t}$}} \quad \boldsymbol{C} \label{GridOp}, \\ 
\textrm{s.t.} ~ 
 \eqref{Market1}-\eqref{BatteryLimit}, \eqref{BatteryCD_linear1}-\eqref{BatteryCD_linear2}, \eqref{Curtail1}-\eqref{VSecurity}.
\end{aligned}
\end{equation}

\section{Cyber Insurance Design}
\label{section:CyberInsurance}

The proposed cyber insurance framework offers a policy for grid operators to manage the financial risks associated with extreme load variations, including those resulting from deliberate manipulation. Unlike approaches such as intrusion detection, which aim to identify or block potential threats, the insurance mechanism focuses on addressing the financial consequences of incidents that fall outside expected operating conditions. This form of protection is intended to work alongside existing defense strategies, particularly in scenarios where such defenses may fail or be bypassed \cite{tatipatri2024comprehensive, zografopoulos2023distributed}. When there are effective defense systems, the frequency of critical failures can be reduced, leading to a lower estimated failure probability and, in turn, a reduced insurance premium. Nevertheless, since defense mechanisms cannot always guarantee complete prevention of all attack scenarios, the proposed insurance serves as an essential safeguard to limit the financial exposure of the operator.

This section presents the design of the proposed cyber insurance framework. We first assess the impact of single and multiple load variations on the system's operational costs. A bi-level optimization model is then proposed to quantify the worst-case impact of such variations. Subsequently, an SMP is employed to estimate the probability of extreme events. Finally, the insurance coverage and premium are derived using VaR and TVaR measures based on the distribution of cost outcomes.

\subsection{Uncertainty of Load Variations}
While the optimization of grid operations relies on the nominal load demand, real-time demand frequently deviates from these expected values~\cite{CaliEle}. Following \cite{tan2024scalable}, we assume that the forecasted load demand is reasonably accurate, with actual demand fluctuating around the predicted value, and there is no spatial or temporal correlated uncertainty. Incorporating spatial-temporal correlations could improve risk assessment fidelity and yield more system-aware insurance premiums and coverage; however, in the absence of sufficiently rich data, such detailed modeling risks overfitting and reduced generality. Besides, we consider that load variation at each individual bus exhibits greater randomness compared to the aggregate load variations in the system. This is due to the individual variations tending to average out or cancel each other when aggregated, resulting in the total system load demand~\cite{sajjad2016definitions}. To model the load variations $P^{var}$, \eqref{Uncertainty1P} and \eqref{Uncertainty1Q} limit the load change on every single bus to the range $\delta$ of the expected load profile. The total load variations in the system at any time period should not exceed $\Delta$ of the sum of expected load demands as stated in \eqref{Uncertainty2P}-\eqref{Uncertainty2Q}:
\begin{equation} \label{Uncertainty1P}
    -\delta P^{Dem}_{i,t} \leq P^{var}_{i,t} \leq \delta P^{Dem}_{i,t},
\end{equation} 
\begin{equation} \label{Uncertainty1Q}
    -\delta Q^{Dem}_{i,t} \leq Q^{var}_{i,t} \leq \delta Q^{Dem}_{i,t},
\end{equation} 
\begin{equation} \label{Uncertainty2P}
    -\Delta\sum_{i\in \mathcal{B}}P^{Dem}_{i,t} \leq \sum_{i\in \mathcal{B}}P^{var}_{i,t} \leq \Delta\sum_{i\in \mathcal{B}}P^{Dem}_{i,t}.
\end{equation}  
\begin{equation} \label{Uncertainty2Q}
    -\Delta\sum_{i\in \mathcal{B}}Q^{Dem}_{i,t} \leq \sum_{i\in \mathcal{B}}Q^{var}_{i,t} \leq \Delta\sum_{i\in \mathcal{B}}Q^{Dem}_{i,t}.
\end{equation}

\begin{algorithm}[t]
\renewcommand{\algorithmicrequire}{\textbf{Input:}}
\renewcommand{\algorithmicensure}{\textbf{Output:}}
\caption{Algorithm for Monte Carlo simulation for sampling operational cost $\boldsymbol{C}$}
\label{alg:algorithm1}
\begin{algorithmic}[1]
    \REQUIRE load profile data
    \ENSURE $S=\{\boldsymbol{C}^*_n\},~ n=1, 2, \dots, N$
    \\ \textit{Initialisation} ${P}^{{Dem}}_{i,t},{Q}^{{Dem}}_{i,t} \leftarrow $ load profile data
    \FOR{$n \leftarrow 1$ to $N$}
        \REPEAT
            \STATE ${P}^{{var}}_{i,t,n},{Q}^{{var}}_{i,t} \leftarrow \{{P}^{{Dem}}_{i,t},{Q}^{{Dem}}_{i,t}\} \cdot X \sim \mathcal{N}(0, (\delta/3)^2)$
        \UNTIL  \small{{$ -\Delta\sum_{i \in B} {P}^{Dem}_{i,t} \leq \sum_{i \in B} P^{var}_{i,t,n} \leq \Delta \sum_{i \in B} {P}^{Dem}_{i,t}$}}\normalsize
        \\ \text{and} \small{{$ -\Delta\sum_{i \in B} {Q}^{{Dem}}_{i,t} \leq \sum_{i \in B} Q^{{var}}_{i,t,n} \leq \Delta \sum_{i \in B} {Q}^{{Dem}}_{i,t}$}}\normalsize
        
        \STATE $\hat{P}^{{Dem}}_{i,t,n} \leftarrow {P}^{{Dem}}_{i,t} + P^{{var}}_{i,t,n}$
        \\$\hat{Q}^{{Dem}}_{i,t,n} \leftarrow Q^{{Dem}}_{i,t} + Q^{{var}}_{i,t,n}$
        
        \STATE Run grid-level optimization \eqref{GridOp} to obtain the optimal operational cost $\boldsymbol{C}_n^*$ under $\hat{P}^{{Dem}}_{i,t,n}, \hat{Q}^{{Dem}}_{i,t,n}$
    \ENDFOR
\end{algorithmic} 
\end{algorithm}

Variations in load demand can lead to different operational costs $\boldsymbol{C}$. 
It is essential to examine the potential operational costs arising from load variation uncertainty within acceptable limits. 
In practice, power systems differ significantly in the types and levels of cybersecurity defenses they employ \cite{WorldBank2022SmartGrids}. This variability motivates the need for an adaptable modeling approach. Accordingly, the proposed insurance framework is designed to be adaptable across a spectrum of scenarios, taking into account the presence and effectiveness of deployed defense mechanisms. By employing a Monte Carlo simulation and adjustable distribution fitting, the framework captures the statistical characteristics of operational risks under different configurations. This allows the resulting risk assessment and insurance pricing to realistically reflect the level of protection embedded in the evaluated system. We employ the Monte Carlo simulation described in Algorithm~\ref{alg:algorithm1} to generate random samples of the load demands $\hat{P}^{Dem}$ and calculate their corresponding optimal operational cost $\boldsymbol{C}^*_n$, denoted as set $S$. The load profile data of the original test system is treated as the nominal load demand ${P}^{Dem}$. Following the work in \cite{tan2024scalable}, we use a Gaussian distribution $\mathcal{N}(0, (\delta/3)^2$ to model the random uncertainties of load variations with a confidence level at $99.7\%$. The uncertain $P^{Dem}$ values are then summed and compared with the total expected load demand $\sum_{i \in \mathcal{B}}{P}^{Dem}_{i,t}$ in the system to ensure they remain within $\Delta$. This approach implies a more predictable total load demand across the entire system compared to the load demand at individual buses. The Monte Carlo simulation result can then be fitted into a distribution function $g(x)$ to represent the probability density function (PDF) of the operational costs. 

In addition to the uncertainty in load demand, attackers might manipulate actual power consumption to harm the grid, thereby bypassing the grid’s detection mechanisms \cite{lakshminarayana2022load}. In this work, we consider that attackers can control the load at each bus within a bounded range through LAAs. A well-designed LAA is able to maximize the operational cost while ensuring that the manipulated load $P^{LAA}$ remains within the uncertainty limits $\delta$ (per-bus) and $\Delta$ (system-wide). In other words, although these load demand variations may appear to be common uncertainties, they are actually being deliberately manipulated to increase the grid's operational costs.

\subsection{Maximum Impact of the Load Variation and LAA}
Since cyber insurance is concerned with the extreme situations of possible load variations or LAAs, it is important for insurers to understand the potential highest impact of load variations and the corresponding maximum operational costs.

Nevertheless, Monte Carlo simulations rely on random sampling, making it difficult to capture exactly the worst-case realization that maximizes the cost. Such scenarios are typically addressed within robust optimization techniques that follow a min-max framework, wherein the objective is to safeguard system performance against all admissible adverse scenarios. However, identifying the specific uncertainty realization that induces the maximum system operation cost requires a max-min approach, such as illustrated in \cite{fang2021data}. In this setting, deliberate load variations act as adversaries that seek to maximize the system operational costs, while the grid operator minimizes them. To capture this interaction, we propose a bi-level problem based on a max-min structure. The bi-level optimization model is presented in \eqref{biH}-\eqref{biL}. The UL problem aims to maximize the operational cost, $\boldsymbol{C'}$, by optimizing the load variations $P^{var}$. (\ref{biH}a)-(\ref{biH}d) constrain the ranges of load variations, which are linear. The LL problem corresponds to the grid optimization discussed in Section~\ref{section:GridModel}. In this context, the load demands $P^{Dem}_{i,t}$ in \eqref{LC},\eqref{Curtail1}-\eqref{Curtail4} are adjusted to $P^{Dem}_{i,t}+P^{var}_{i,t}$ to account for the load variations. The LL objective is convex, and all system constraints, including those for energy transactions, BESS, load curtailment, and load variations, are linear or convex, allowing the optimization problem to be solved efficiently. 

\small{
\begin{flalign}
\mathop {maximize}\limits_{\textbf{$P^{var}_{i,t},Q^{var}_{i,t}$}} & \qquad
\boldsymbol{C'}  \label{biH}\\
\textrm{s.t.} \qquad 
& -\delta   {P}^{Dem}_{i,t} \leq P^{var}_{i,t} \leq \delta   {P}^{Dem}_{i,t},  \tag{\ref{biH}{a}} \\
& -\delta   {Q}^{Dem}_{i,t} \leq Q^{var}_{i,t} \leq \delta   {Q}^{Dem}_{i,t},  \tag{\ref{biH}{b}}
\allowdisplaybreaks[4]\\
&-\Delta\sum_{i \in \mathcal{B}}{P}^{Dem}_{i,t} \leq \sum_{i \in \mathcal{B}}P^{var}_{i,t} \leq \Delta\sum_{i \in \mathcal{B}}{P}^{Dem}_{i,t}, \tag{\ref{biH}{c}}\\
&-\Delta\sum_{i \in \mathcal{B}}{Q}^{Dem}_{i,t} \leq \sum_{i \in \mathcal{B}}Q^{var}_{i,t} \leq \Delta\sum_{i \in \mathcal{B}}{Q}^{Dem}_{i,t}, \tag{\ref{biH}{d}}\\
&\mathop{minimize}\limits_{\textbf{$P^{Buy}_{i,t}$, $P^{Sell}_{i,t}$,$P^{Curtail}_{i,t}$,$Q^{PV}_{i,t}$}} \quad \boldsymbol{C} \label{biL}\\
&\textrm{s.t.} \quad\eqref{Market1}-\eqref{BatteryRemain}, \eqref{BatteryCD_linear1}-\eqref{BatteryCD_linear2}, \eqref{PV}-\eqref{VSecurity}, \tag{\ref{biL}{a}} \\
&\qquad\quad P^{Load}_{i,t} = {P}^{Dem}_{i,t} + P^{var}_{i,t} - P^{Curtail}_{i,t}, \tag{\ref{biL}{b}}\\
&\qquad\quad Q^{Load}_{i,t} = {Q}^{Dem}_{i,t} + Q^{var}_{i,t} - Q^{Curtail}_{i,t}, \tag{\ref{biL}{c}}\\
&\qquad\quad 0 \leq P^{Curtail}_{i,t} \leq \gamma({P}^{Dem}_{i,t}+P^{var}_{i,t}),\tag{\ref{biL}{d}}\\
&\qquad\quad -\gamma({Q}^{Dem}_{i,t}+Q^{var}_{i,t}) \leq Q^{Curtail}_{i,t} \notag\\
&\qquad\qquad\qquad \leq \gamma(\hat{Q}^{Dem}_{i,t}+Q^{var}_{i,t}),  \tag{\ref{biL}{e}}
\end{flalign}
}
\normalsize

The solution of this bi-level optimization problem is then incorporated into the samples derived by the Monte Carlo simulation, in Algorithm \ref{alg:algorithm1}, for a more accurate fitting of the distribution function of system operational costs. The effectiveness of this consideration is demonstrated explicitly in the simulation results section.

\subsection{Risk Probability Estimation} \label{subsection:SMP}
While the bi-level optimization can determine the maximum impact of the load variation $\boldsymbol{C'}$, it is also crucial to estimate the likelihood of such extremes to evaluate the risks associated with the attack.
An SMP framework can determine the probability of an attack using minimal prior information, offering a flexible and adaptive framework for decision-making in dynamic and uncertain environments~\cite{lau2022novel}. In this work, we utilize the SMP model in~\cite{acharya2021cyber}. This general model evaluates the probability of cyber-attacks and enables the calculation of the probability $\boldsymbol{P}$ of extreme events considered by cyber insurance.
In this model, the states of the system are defined as: \textbf{(1) Good}: The system is not exposed to the risks of extreme situations or cyber-attacks, and it is operating in an optimal way; 
\textbf{(2) Vulnerability}:
The system is suffering from abnormal actions, and an attack is considered to be launched; 
\textbf{(3) Detection}: The system has detected that there is an attack. The corresponding alarm mechanism is triggered; 
\textbf{(4) Containment}: Due to the detection of attacks, defense mechanisms are activated to restore the system to a good state; 
\textbf{(5) Failure}: The attack is not detected, and no defense mechanism is triggered. The system fails. 

\begin{figure}[t]
    \centering
    \includegraphics[width=0.3\textwidth]{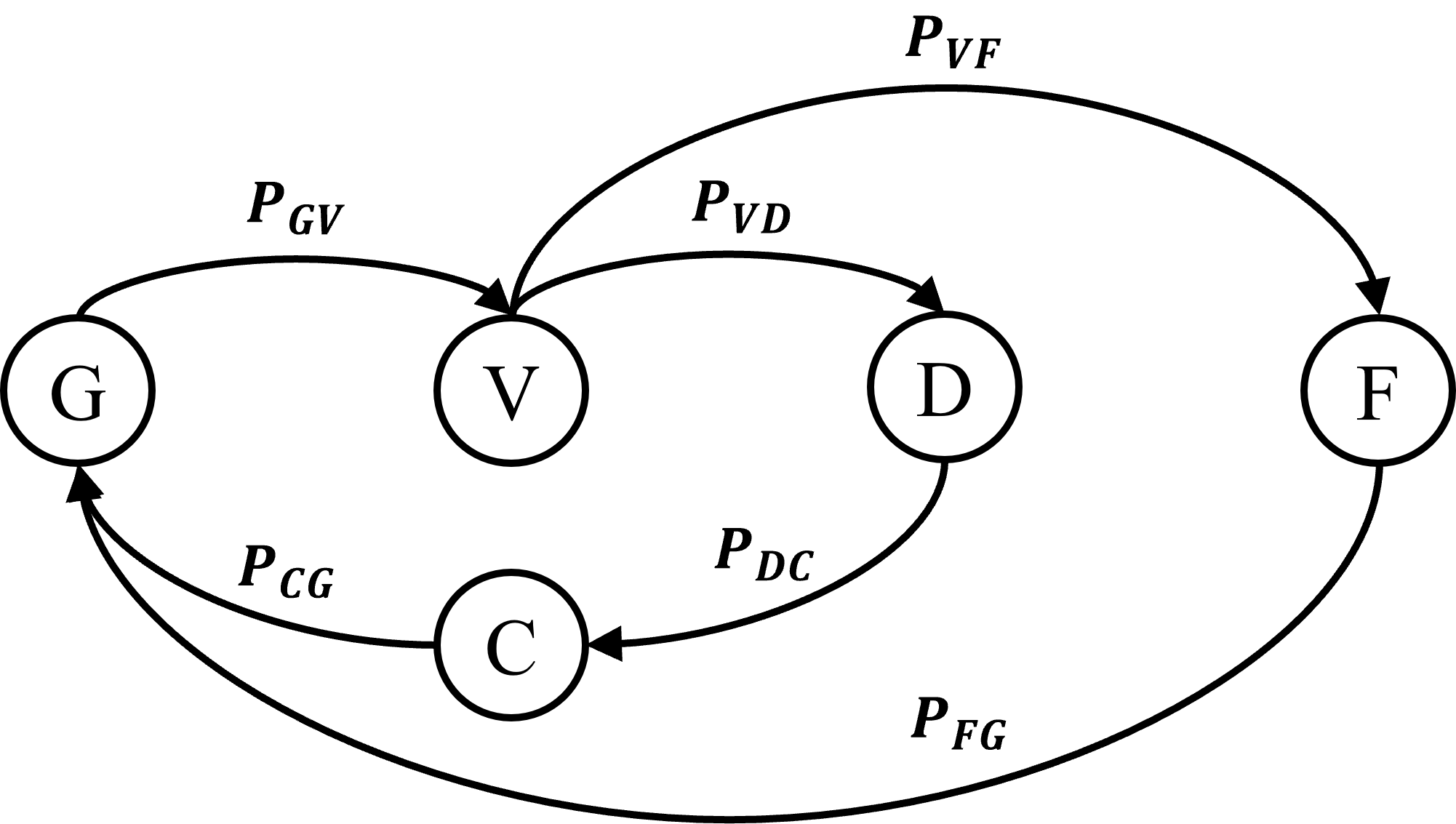}
    \caption{Semi-Markov process (SMP) model for risk probability estimation. P(·) defines the CDF of the state transition time.}
    \label{fig:SMP}
\end{figure}

The states in the designed SMP model are shown in Fig.~\ref{fig:SMP}. When an attacker initiates an attack on the system, the state transitions from \textbf{Good} to \textbf{Vulnerability}. If the system detects and identifies the attack, the state moves from \textbf{Vulnerability} to \textbf{Detection}. Once the attack is identified and mitigation mechanisms are activated, the state transitions from \textbf{Detection} to \textbf{Containment}. Upon successful recovery from the attack, the state returns to \textbf{Good}. However, if the attack is not successfully detected in the \textbf{Vulnerability} state, the system transitions to \textbf{Failure}. After experiencing a \textbf{Failure}, the system eventually returns to its normal operational state, \textbf{Good}. 

The proposed SMP enables the consideration of various factors that can practically influence the system operation. For instance, the model captures the likelihood of LAAs occurring and targeting the power system, where higher likelihoods correspond to a greater probability of transitioning from the \textbf{Good} to the \textbf{Vulnerability} state, i.e., a higher $P_{GV}$. The probability estimation scheme also incorporates the impact of defense mechanisms designed to protect the power system against LAAs. Specifically, when effective defense measures are implemented, the probability of transitioning from the \textbf{Vulnerability} to the \textbf{Failure} state ($P_{VF}$) decreases, while the probability of transitioning to the \textbf{Detection} state ($P_{VD}$) increases. By systematically adjusting the transition relationships among the SMP states, the proposed framework enables a comprehensive analysis of the economic impact of LAAs under various power system infrastructures and operating conditions.

The kernel matrix of the SMP can be expressed as follows: 

\small{
\begin{equation} \label{Kernel}
    \boldsymbol{K}(\tau) = \begin{bmatrix}
0 & k_{GV}(\tau) & 0 & 0 & 0\\
0 & 0 & k_{VD}(\tau) & 0 & k_{VF}(\tau)\\
0 & 0 & 0 & k_{DC}(\tau) & 0\\
k_{CG}(\tau) & 0 & 0 & 0 & 0\\
k_{FG}(\tau) & 0 & 0 & 0 & 0
    \end{bmatrix},
\end{equation}}\normalsize
where $k_{mn}(\tau)$ is the probability that the SMP has just entered state $m$, and will transition to the next state $n$ within time $\tau$. Each time step of $\tau$ corresponds to the timestamp $t$ for load variations, i.e., state transitions can potentially occur hourly. 

We employ Weibull distributions to represent the state transition times, as this distribution function is commonly used in failure analysis~\cite{acharya2021cyber}. The cumulative distribution function (CDF) $F(\tau,\lambda,\beta)$ of a Weibull distribution is given below:
\begin{equation} \label{CDF_Weibull}
    F(\tau,\lambda,\beta) = 1-e^{-(\frac{\tau}{\lambda})^\beta},
\end{equation} 
where $\lambda$ is the scale parameter and $\beta$ is the shape parameter. Consequently, the $k_{mn}(\tau)$ can be determined as:

\small{
\begin{equation} \label{Kernel_Weibull_k}
\begin{aligned}
    &k_{GV}(\tau) = F_{GV}(\tau,\lambda_{GV},\beta_{GV}),\\
    &k_{VD}(\tau) = \int^\tau_0\overline{F}_{VF}(\tau,\lambda_{VF},\beta_{VF})d F_{VD}(\tau,\lambda_{VD},\beta_{VD}),\\
    &k_{VF}(\tau) = \int^\tau_0\overline{F}_{VD}(\tau,\lambda_{VD},\beta_{VD})d F_{VF}(\tau,\lambda_{VF},\beta_{VF}),\\
    &k_{DC}(\tau) = F_{DC}(\tau,\lambda_{DC},\beta_{DC}),\\
    &k_{CG}(\tau) = F_{CG}(\tau,\lambda_{CG},\beta_{CG}),\\
    &k_{FG}(\tau) = F_{FG}(\tau,\lambda_{FG},\beta_{FG}),
\end{aligned}
\end{equation}}\normalsize 
where $\overline{F}=1-F$. Note that $k_{VD}(\tau)$ and $k_{VF}(\tau)$ are not intuitively the CDF of the corresponding state transitions. This is because state \textbf{Vulnerability} has two potential destinations, \textbf{Detection} and \textbf{Failure}, and thus their calculations involve the dependencies between the two transitions. 

The steady-state transition probability matrix of SMP $\boldsymbol{K}(\infty)$ can be represented as: 

\small{
\begin{equation}
 \label{Kernel_inf}
    \begin{bmatrix}
0 & k_{GV}(\infty) & 0 & 0 & 0\\
0 & 0 & k_{VD}(\infty) & 0 & k_{VF}(\infty)\\
0 & 0 & 0 & k_{DC}(\infty) & 0\\
k_{CG}(\infty) & 0 & 0 & 0 & 0\\
k_{FG}(\infty) & 0 & 0 & 0 & 0
    \end{bmatrix}. 
\end{equation}}\normalsize

Solving the embedded Markov chain (EMC) steady-state equations gives the steady-state probabilities $\boldsymbol{v}$ for the SMP:
\begin{equation} \label{EMC1}
    \boldsymbol{v} = \boldsymbol{v} \boldsymbol{K}(\infty),
\end{equation} 
\begin{equation} \label{EMC2}
    \boldsymbol{v} \boldsymbol{e}^\top = 1.
\end{equation} 
where $\boldsymbol{v} = [v_G~v_V~v_D~v_C~v_F]$ and $\boldsymbol{e} = [1~1~1~1~1]$.

To calculate the sojourn time $T_n$ in state $n$ before transitioning, we refer to Kumar \textit{et al.}~\cite{kumar2013availability}:

\small{
\begin{equation} \label{Kernel_Weibull_T}
\begin{aligned}
    &T_G = \int^\infty_0 \overline{F}_{GV}(\tau,\lambda_{GV},\beta_{GV})~d\tau,\\
    &T_V = \int^\infty_0 \overline{F}_{VD}(\tau,\lambda_{VD},\beta_{VD}) \overline{f}_{VF}(\tau,\lambda_{VF},\beta_{VF})~d\tau,\\
    &T_D = \int^\infty_0 \overline{F}_{DC}(\tau,\lambda_{DC},\beta_{DC})~d\tau,\\
    &T_C = \int^\infty_0 \overline{F}_{CG}(\tau,\lambda_{CG},\beta_{CG})~d\tau,\\
    &T_F = \int^\infty_0 \overline{F}_{FG}(\tau,\lambda_{FG},\beta_{FG})~d\tau.
\end{aligned}
\end{equation}} \normalsize

Incorporating the steady-state probabilities of EMC solved by \eqref{EMC1}-\eqref{EMC2}, the steady-state probability of
transitioning to a certain state $n$ can be obtained by:
\begin{equation} \label{Probability}
    P_n = \frac{v_n T_n}{\boldsymbol{v}\boldsymbol{T}^\top},
\end{equation} 
where $\boldsymbol{T} = [T_G~T_V~T_D~T_C~T_F]$. $P_F$ represents the steady-state probability of a failure state. Its value reflects the likelihood of undetected cyber-attacks or extreme load variations leading to substantial operational costs.

\subsection{Insurance Policy and Premium}
In this subsection, we design the cyber insurance policy and determine the corresponding premium based on the risk measures using VaR and TVaR. Specifically, we utilize the failure probability $P_F$ as the confidence level $\alpha$ in the subsequent financial risk evaluation.

To calculate the risk measures in the context of power system operations, we first characterize the distribution of optimal operational costs resulting from different load variations. The VaR and TVaR metrics \cite{lau2021coalitional} are derived from the probability distribution of the optimal operational cost $C^*$, which is obtained by solving the grid-level optimization problem \eqref{GridOp} under a wide range of stochastic and adversarial load variation scenarios generated via Monte Carlo simulation and bi-level optimization. This cost distribution $g(x)$ reflects the system’s sensitivity to load uncertainty and forms the basis for financial risk evaluation.

VaR at confidence level $\alpha$ is defined as the minimum cost level beyond which only $(1 - \alpha) \times 100\%$ of the most extreme cost outcomes occur:

\begin{equation} \label{VaR}
     VaR_\alpha= \{x\in \mathbb {R} :G(x)>\alpha\},
\end{equation}
where $G(x)$ is the CDF of $g(x)$. In the insurance policy, $VaR_\alpha$ serves as a coverage parameter, i.e., it sets the loss threshold beyond which insurance compensation becomes active.

TVaR complements VaR by capturing the expected severity of losses that exceed its threshold. It is computed as:
\begin{equation} \label{TVaR}
    TVaR_\alpha = \frac{1}{1-\alpha}\int^{\infty}_{VaR_\alpha}(x-1)g(x)dx.
\end{equation}It provides a risk estimation that accounts for the worst-case financial losses within the distribution tail. In our framework, the insurance premium that the operator pays equals:

\begin{equation} \label{Premium}
    Premium_\alpha= \alpha\cdot TVaR_\alpha,
\end{equation}
which is a risk-adjusted pricing mechanism that accounts for both the likelihood $\alpha$ and severity $TVaR_\alpha$ of extreme operational costs.

\begin{figure}[t]
    \centering    
    \includegraphics[width=0.45\textwidth]{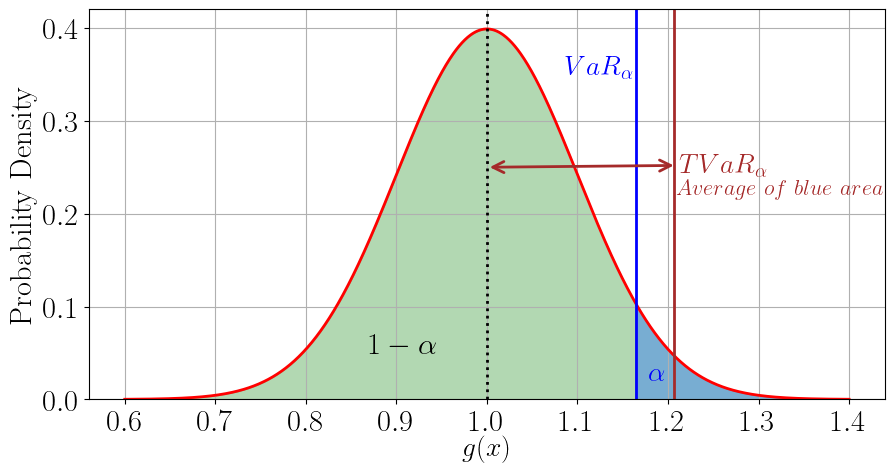}
    \caption{Illustration of the operational cost distribution $g(x)$, highlighting $VaR_\alpha$ as the insurance activation threshold, and $TVaR_\alpha$ as the expected loss in the tail region (blue area).}
    \label{fig:VaRTVaR}
\end{figure}

Consequently, the integration of VaR and TVaR into the proposed framework establishes a quantitative link between the power system’s operational risk and the economic terms of the cyber insurance policy. An illustration of the $VaR_\alpha$ and $TVaR_\alpha$ metrics is provided in Fig.~\ref{fig:VaRTVaR}, where the blue tail region denotes insured losses beyond the activation VaR threshold. In short, with the VaR and TVaR, an insurance policy can be designed such that if the grid operator purchases a premium with $Premium_\alpha$, the insurer will cover the operational costs exceeding $VaR_\alpha$.

\begin{figure}[t]
    \centering    \includegraphics[width=0.4\textwidth]{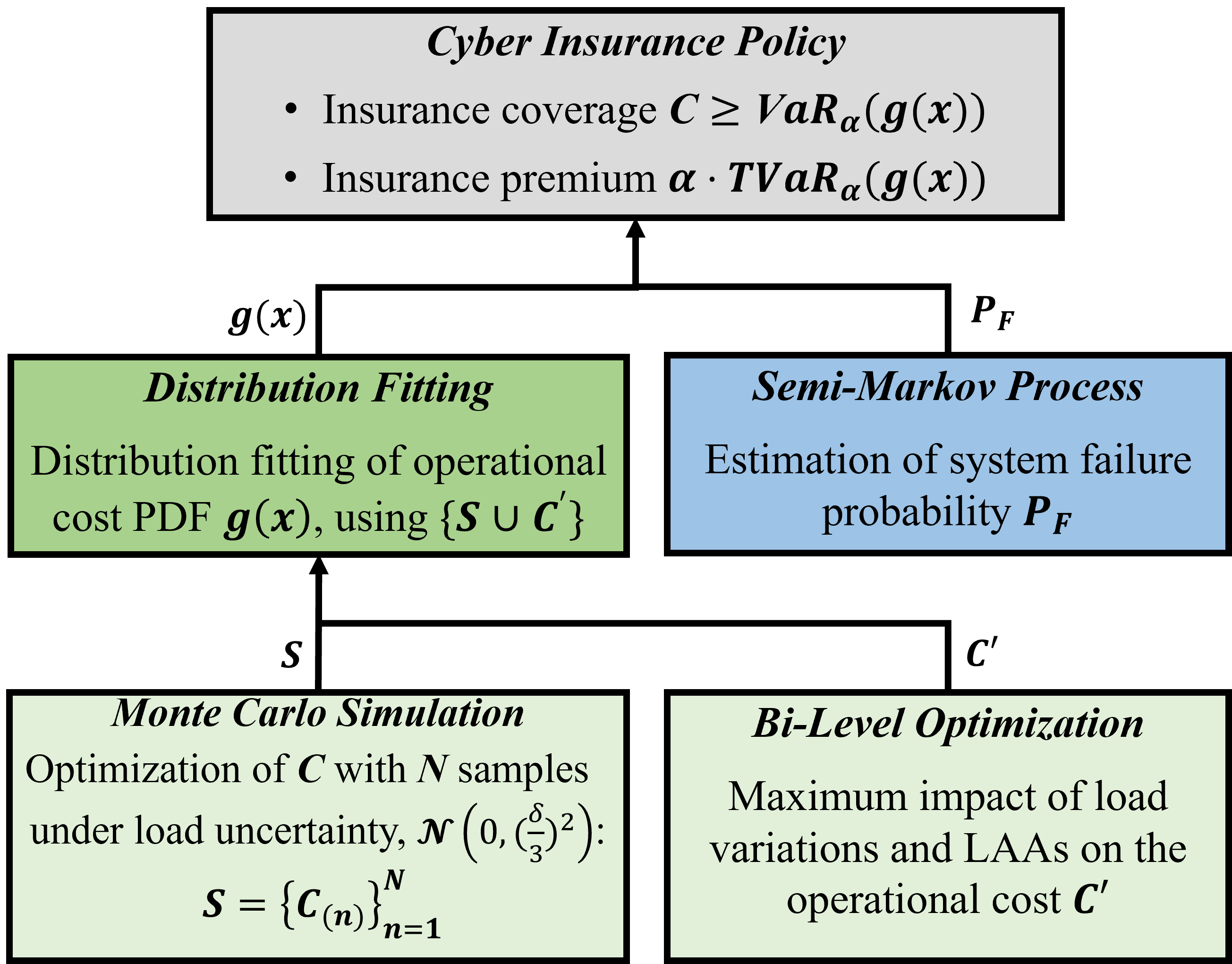}
    \caption{Detailed insurance framework.}
    \label{fig:detail}
\end{figure}

The detailed steps of the cyber insurance design are shown in Fig.~\ref{fig:detail}. First, Monte Carlo simulations and a bi-level optimization are used to determine the grid's operational cost behavior under different load variations. This behavior is then fitted into a distribution function to analyze the probability distribution and identify potential risks and cost implications. Meanwhile, an SMP is designed to estimate the probability of the conditions considered in the insurance policy. Finally, the insurance coverage and premium are calculated based on VaR and TVaR.

\section{Simulation Results}
\label{section:Simulation}
This section presents two case studies that demonstrate the effectiveness of the proposed cyber insurance policy in safeguarding grid operators against additional operation costs arising from extreme load variations and potential LAAs. The first case study considers a modified IEEE 118-bus system \cite{zhang2007improved}, while the second one demonstrates the applicability of the proposed cyber insurance framework on the larger IEEE European low-voltage test feeder \cite{IEEE_European_LV_Test_Feeder_v2}. All experiments were carried out on a computer with an AMD Ryzen 9 9950X @ 4.29GHz and 16 GB of RAM. The proposed model is implemented in Pyomo 6.5.0 with Python 3.10.15, and solved by IBM ILOG CPLEX 22.1.1. Table \ref{Table:Runtime} summarizes the runtimes obtained for the considered case studies, which remain within an acceptable range since the optimization is performed for computing the optimal insurance premium and coverage. The modified IEEE 118-bus system includes $7$ PV plants and $7$ BESS, whose locations and capacities are designed based on \cite{huy2023optimal}.
The capacities of the PVs and BESS are provided in Table~\ref{Table:Capacity}. The other system parameters are the same as the setup in \cite{zhang2007improved}. 

\begin{table}[t]
    \centering
    \caption{Average runtimes for model compiling and solving}
    \begin{tabular}{ccc}
    \toprule
    System & Compile (s) & Solve (s) \\
    \midrule
    IEEE 906-bus & 18.71 & 9.94 \\
    IEEE 118-bus & 1.26 & 7.09 \\
    \bottomrule
    \label{Table:Runtime}
    \end{tabular}
\end{table}

\begin{table}[t]
    \centering
    \caption{The Capacity of PVs and BESS in the System (MVA)}
    \begin{tabular}{cccccccc} 
    \toprule 
     \textbf{PV} & \#30 & \#50 & \#58 & \#74 & \#80 & \#96 & \#110 \\
     \textbf{Capacity} & 5.0  & 3.0  & 2.0  & 3.0 & 3.0  & 3.0  & 4.0   \\ 
    \midrule
     \textbf{BESS} & \#30 & \#50 & \#58 & \#74 & \#80 & \#96 & \#110 \\
     \textbf{Capacity} & 2.0  & 2.0  & 2.0  & 2.0 & 2.0  & 2.0  & 2.0 \\ 
    \bottomrule 
    \label{Table:Capacity}
    \end{tabular}
\end{table}

\begin{figure}[t]
    \centering
    \subfigure[]{
        \includegraphics[width=0.225\textwidth]{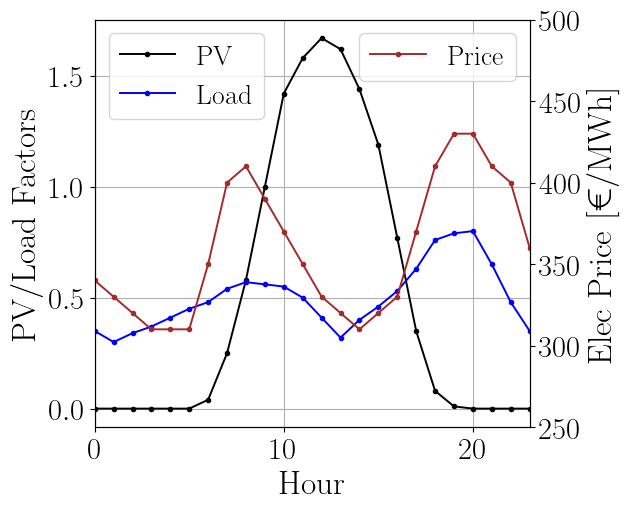}
        \label{fig:PVLoadPrice}
    }
    \hfill
    \subfigure[]{
        \includegraphics[width=0.225\textwidth]{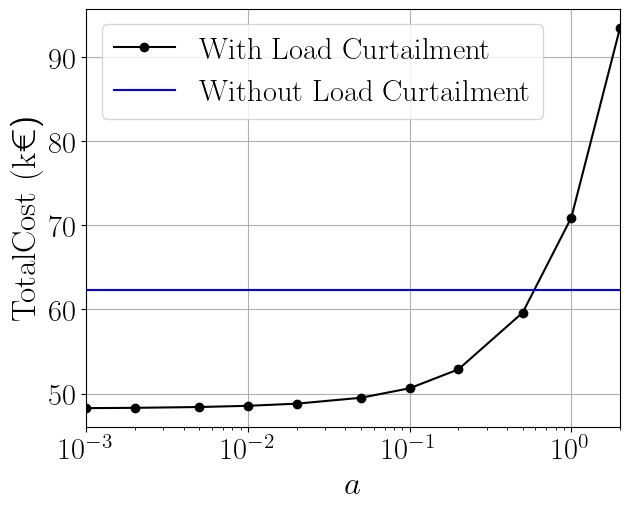}
        \label{fig:LCCost}
    }
    \subfigure[]{
        \includegraphics[width=0.225\textwidth]{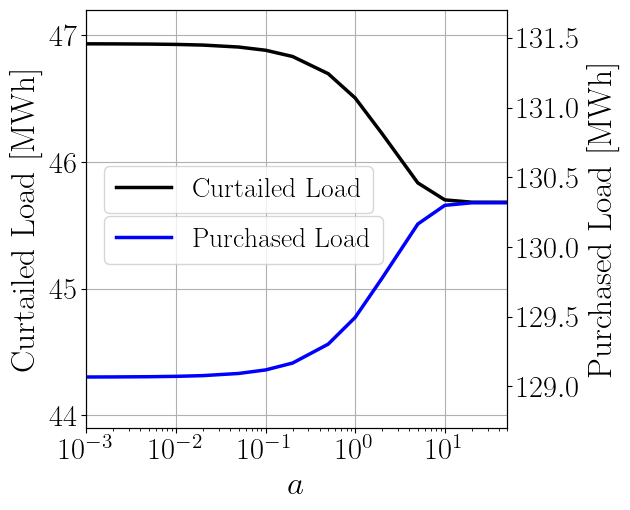}
        \label{fig:CurtialMarket}
    }
    \hfill
    \subfigure[]{
        \includegraphics[width=0.225\textwidth]{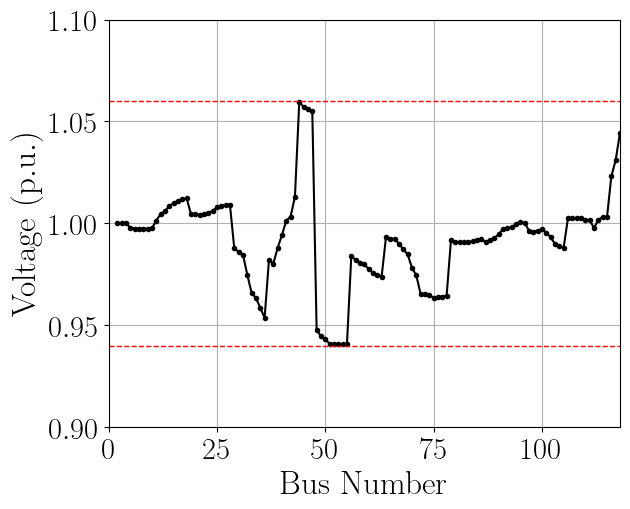}
        \label{fig:Voltages}
    }
    \caption{(a) PV, load factors and electricity prices over 24h, (b) Total operational cost of the modified IEEE 118-bus system using the cost function $\boldsymbol{c}^{LC}$ with different $a$, and $b=0$, (c) Total curtailed loads and net purchased electricity of a day in the modified IEEE 118-bus system for different $a$, (d) Bus voltages of the modified IEEE 118-bus system after optimization with $a=1$.}
    \label{fig:IEEE118_combined}
\end{figure}

The load profile of each bus is the same as the one in the original IEEE 118-bus system, but with load factors over 24 hours. The power generated by PVs also has an hourly-changed factor representing its variation during the day. These factors are mapped through publicly available data \cite{electricitymaps2025} and are given together with the electricity prices \cite{gridX} in Fig. \ref{fig:PVLoadPrice}.

We first show the behavior of operational cost and load curtailment with the proposed optimization model. Fig.~\ref{fig:LCCost} shows the optimal cost $\boldsymbol{C}$ under different curtailment penalty $a$. It can be observed that the cost increases with $a$ and eventually exceeds the total cost without load curtailment. This occurs because the optimization problem is cost-driven, selecting the approach (curtailing the load or buying electricity) that has the lower cost. Therefore, the optimal solution is to do more curtailment if the cost of load curtailment compensation is lower than the price of purchasing electricity from the market. On the other hand, $b$ is formulated as a constant in the insurance framework, representing the one-time incentive, and does not affect the solution of the optimization problem. As illustrated in Fig. \ref{fig:CurtialMarket}, in the beginning, the cost of load curtailment was low; the system prioritized curtailing the load over buying electricity from the energy market. When $a$ becomes higher than $0.5$, it costs less to buy electricity from the markets, and thus, more electricity is bought. However, due to the system's voltage security constraint \eqref{VSecurity}, a specific amount of loads still needs to be curtailed to maintain the bus voltages within acceptable margins. As shown in Fig. \ref{fig:Voltages}, the buses around bus $50$ experience significantly low voltages, which touch the limit $0.94$ p.u. even when the load demands are curtailed. If there is no load curtailment, the voltages will drop even lower, which violates the security limits. In other words, a certain level of curtailment remains necessary, and the increasing value of $a$ thus leads to a linear growth of the total cost as shown in the right side of Fig.~\ref{fig:LCCost}. 

Table~\ref{Table:ab} illustrates how different values of the curtailment penalty parameter $a$ influence the compensation $b$ offered to users. As discussed in Section~\ref{section:GridModel}, given that curtailed users may experience utility loss, $b$ encourages participation by compensating each user for the savings their load curtailment brings to the system. Though the system's cost is not directly reduced from load curtailment, the grid operator gains the flexibility to curtail the load under emergencies, thereby enhancing the system’s ability to address significant supply and demand discrepancies. When $a$ becomes excessively large, the system cost increases to the point where load curtailment no longer yields savings (as shown in Fig.~\ref{fig:LCCost} where the black curve rises above the blue curve). In such cases, the computed value of $b$ turns even negative (denoted with '$\times$' in Table \ref{Table:ab}), implying that users would need to pay the operator to participate in curtailment, which is impractical. Therefore, $a$ should be selected within a range that ensures curtailment remains economically beneficial.

\begin{table}[t]
    \centering
    \caption{Different Values of Parameter $a$ and $b$ in $\boldsymbol{c}(LC)$}
    \begin{tabular}{cc|cc|cc} 
    \bottomrule \rule{0pt}{8pt}
    $a$ & $b$ & $a$ & $b$ & $a$ & $b$ \\  
    \hline \rule{0pt}{8pt}
    0.001 & 14.062& 0.05& 12.829 & 2& × \\ 
    0.002 & 14.026& 0.1& 11.699 & 5& × \\
    0.005 & 13.929& 0.2& 9.452 & 10& × \\
    0.01 & 13.785& 0.5& 2.7 & 20& × \\
    0.02 & 13.529& 1& × & 50& × \\
    \toprule 
    \label{Table:ab}
    \end{tabular}
\end{table}

\begin{figure}[t]
    \centering
    \subfigure[]{
        \includegraphics[width=0.4\textwidth]{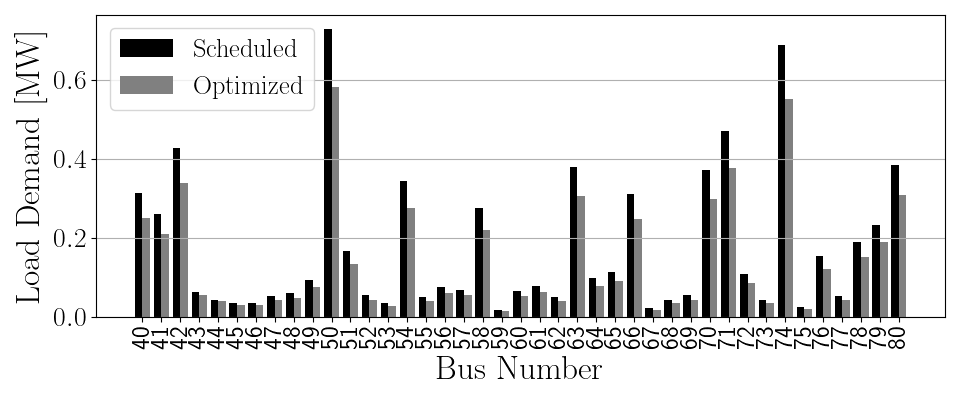}
        \label{fig:8PM}
    }
    \subfigure[]{
        \includegraphics[width=0.4\textwidth]{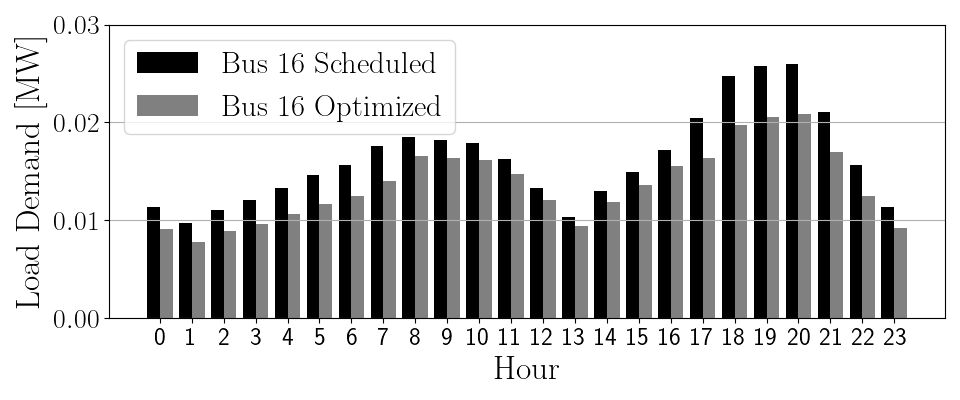}
        \label{fig:Bus16}
    }
    \caption{(a) Load curtailment in the 118-bus system at $8$ p.m. (Bus 40 to Bus 80), (b) Load curtailment on bus $16$ over 24 hours of the day.}
    \label{fig:LC_combined}
\end{figure}

Fig.~\ref{fig:8PM} shows a part of the load curtailments (buses $40$-$80$) in the 118-bus system at $8$ p.m. It can be observed that buses with higher load levels have more significant curtailment during peak hours, as the optimized values are noticeably lower than the scheduled ones. This is because of the load curtailment compensation function \eqref{LC}. When both buses need to curtail a certain amount of load, the bus with higher demand will experience a smaller utility loss, leading to lower compensation for load curtailment. As a result, the system prioritizes curtailing larger loads first. Fig.~\ref{fig:Bus16} shows the amount of curtailed load on a specific bus (bus $16$) in a day. From $5$ p.m. to $9$ p.m., the loads are curtailed significantly. This is because the load demand peaks during this period, while electricity prices also have the highest values. The shortages in the PV generation also bring the necessity to curtail more loads to balance the supply and demand.

\subsection{Impact of Load-Altering Attacks (LAAs)}

\begin{figure}[t]
    \centering    \includegraphics[width=0.4\textwidth]{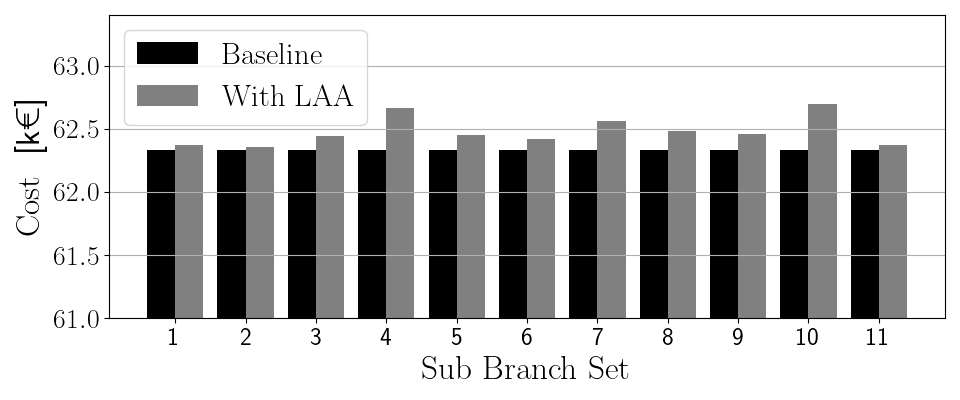}
    \caption{System's daily total operational costs with and without a LAA at 8 p.m. with different victim sub-branches, $a=0.5$.}
    \label{fig:LAAGerman}
\end{figure}

In LAA scenarios, attackers have the capability to compromise a massive number of devices in the power grid to maliciously increase or decrease the load demands on multiple buses~\cite{lakshminarayana2022load}, potentially affecting the operational costs and causing system-wide impacts. We examine the proposed model with increased loads on specific buses. We split the grid into 11 sub-branches (shown in Table \ref{Table:subBranch}), assuming the attacker can manipulate all the load buses on one of the branches. Noting that the highest tariff occurs from 7 p.m. to 9 p.m., coinciding with peak load demands. The loads on the victim sub-branches are maliciously increased by $30\%$ at 8 p.m. It can be observed in Fig.~\ref{fig:LAAGerman} that the system's operational costs increase due to the LAA. The impacts are more significant when performing LAAs on branches with higher load demands. However, due to the flexibility of doing load curtailment and energy transactions in the markets, the overall impact is marginal.  

\begin{table}[t]
    \centering
    \caption{The considered sub-branches of the IEEE 118-bus system.}
    \begin{tabular}{cc} 
    \toprule 
    \textbf{Set Index} & \textbf{Buses} \\  
    \midrule
    1 & \{4,5,6,7,8,9\} \\ 
    2 & \{12,13,14,15,16,17\} \\ 
    3 & \{18,19,20,21,22,23,24,25,26,27\} \\ 
    4 & \{30,31,32,33,34,35,36,37,38,39,40,41,42,43,61,62\} \\ 
    5 & \{44,45,46,47,48,49,50,51,52\} \\ 
    6 & \{53,54,55,56,57,58,59,60\} \\ 
    7 & \{66,67,68,69,70,71,72,73,74,75,76,77\} \\ 
    8 & \{79,80,81,82,83,84,85,86,87,88\} \\ 
    9 & \{89,90,91,92,93,94,95,96,97,98,99\} \\ 
    10 & \{101,102,103,104,105,106,107,108,109,110,112,113,111\} \\ 
    11 & \{114,115,116,117,118\} \\ 
    \bottomrule 
    \label{Table:subBranch}
    \end{tabular}
\end{table} 

\subsection{Impact of Load Variation Uncertainties}

\begin{figure}[t]
    \centering
    \subfigure[]{
        \includegraphics[width=0.4\textwidth]{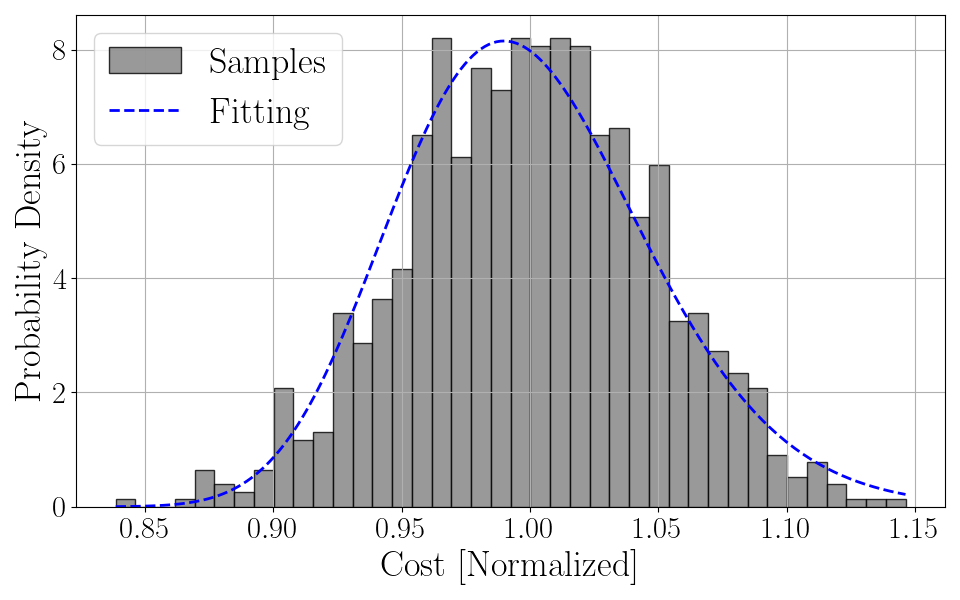}
        \label{fig:Fitting}
    }
    \hfill
    \subfigure[]{
        \includegraphics[width=0.4\textwidth]{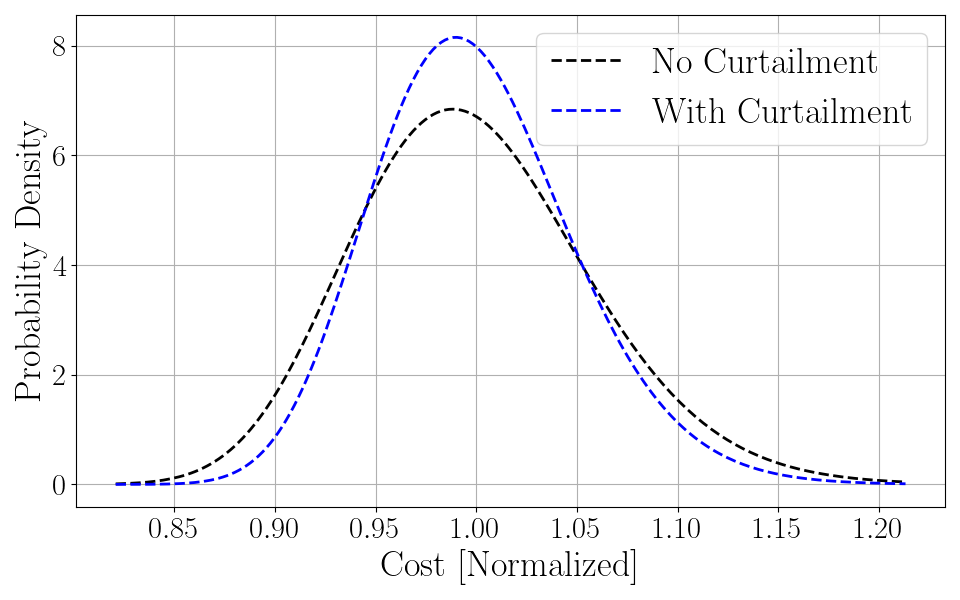}
        \label{fig:Fitting2}
    }
    \caption{(a) PDF of the optimal operational cost, (b) Optimal operational cost with and without load curtailment, $\delta=30\%$, $\Delta=10\%$, $a=0.5$.}
    \label{fig:Fitting_combined}
\end{figure}

As mentioned in Section~\ref{section:CyberInsurance}, the uncertainties of load variations can lead to potentially high operational costs. 
As indicated in~\cite{proedrou2021comprehensive}, in the 2020s, the deviation between the standard load profile and actual consumption, originally expected to be $10$-$20\%$ before, is likely to increase higher due to the widespread use of multiple electronic devices. According to the study in~\cite{stoyanova2013characterization}, the detected load profile deviations for residential buildings in Austria can already surpass $30\%$. In this work, we set the maximum load variation range to $\delta=30\%$ and $\Delta=10\%$. We use the Monte Carlo simulation with $1000$ samples for the optimization problem, with compensation parameter $a=0.5$. As shown in Fig.~\ref{fig:Fitting}, the distribution of the optimal operational cost can be fitted into distribution functions. In this work, we select the inverse Gaussian PDF, which is suitable for asymmetric distributions and best fits the distribution of operational costs.

To assess the impact of load curtailment on the grid's operational costs, we conduct another simulation excluding load curtailments. It can be observed from Fig.~\ref{fig:Fitting2} that the distribution is more concentrated at a lower value with the load curtailment. At the same time, the probability of increased operating costs is lower without load curtailment. This is because load curtailments and participation in the markets enhance the system's flexibility during extreme scenarios, which illustrates the necessity of introducing load curtailment.

Then, we solve the bi-level optimization in (\ref{biH})--(\ref{biL}). Results are presented in Table \ref{Table:Max}. The expected cost of daily operation without considering real-time load variations is \euro$62.33k$. However, with a 10\% variation in the total load demand, this operational cost can be increased even to almost twice the expected. When the tolerance on a single load increases to $50\%$, and the total tolerance is $20\%$, the maximum operational cost can become even higher to three times the expected cost. This emphasizes the need for risk management.

\begin{table}[t]
    \centering
    \caption{The Maximum Daily Operational Cost Due to Load Variation in the System ($k$\euro)}
    \begin{tabular}{ccc} 
    \toprule 
    \quad$\delta=0,\Delta=0$\quad & \quad$\delta=30\%,\Delta=10\%$\quad & \quad$\delta=50\%,\Delta=20\%$\quad \\  
    \midrule
    $62.33$ & $ 73.35 (+17.68\%)$ & $83.13 (+33.37\%)$   \\ 
    \bottomrule 
    \label{Table:Max}
    \end{tabular}
\end{table}

\subsection{Cyber Insurance}
As specified in \eqref{TVaR}, calculating the insurance premium requires defining the confidence level $\alpha$ and the PDF of the operational cost. $\alpha$ represents the probability of an extreme situation occurring in the grid, which can be determined by running the SMP model shown in Fig.~\ref{fig:SMP}. Existing works have provided Weibull distribution models for specific state transitions by fitting public historical data from the information technology (IT) industry available from SANS Institute \cite{ID}, the Canadian Institute for Cybersecurity \cite{GI}, and empirical studies~\cite{franke2014distribution,acharya2021cyber}. These models provide the evaluation of probabilities for the general cyber-attacks, which include the LAAs. Due to the limited public data on the LAAs, we refer to the Weibull CDF provided in those studies to model the SMP of LAAs. The parameters are listed in Table \ref{Table:WBCDF}. The calculated probability of failure is $P_F=3.98\%$, which corresponds to $\alpha=3.98\%$. This value indicates the likelihood of experiencing an extreme load variation within an hour during the day.

\begin{table}[h]
    \centering
    \caption{Weibull CDF Parameters of Each State Transition \cite{acharya2021cyber}}
    \begin{tabular}{cccc} 
    \toprule 
    State Transition & $\lambda$ &  $\beta$ & Reference \\  
    \midrule 
    G-V & 2.0675 & 18.8178 & \cite{GI}\\
    V-D & 1.9293 & 16.0712 & \cite{ID}\\
    D-C & 1.5698 & 18.4858 & \cite{ID}\\
    C-G & 1.3816 & 15.7033 & \cite{ID}\\
    V-F & 0.7000 & 400.000 & \cite{acharya2021cyber}\\
    F-G & 0.6783 & 13.4487 & \cite{franke2014distribution}\\
    \bottomrule 
    \label{Table:WBCDF}
    \end{tabular}
\end{table}

\begin{table}[!ht]
    \centering
    \caption{Inverse Gaussian PDF Distribution Fitting Parameters}
    \begin{tabular}{ccc} 
    \toprule 
    Samples & $\mu$ &  $\lambda$ \\  
    \midrule
    Monte Carlo simulation & 0.9994 & 22.5967 \\
    + bi-level maximum impact & 0.9999 & 31.4798 \\
    \bottomrule 
    \label{Table:fitting}
    \end{tabular}
\end{table}

\begin{figure}[!ht]
    \centering   
    \includegraphics[width=0.4\textwidth]{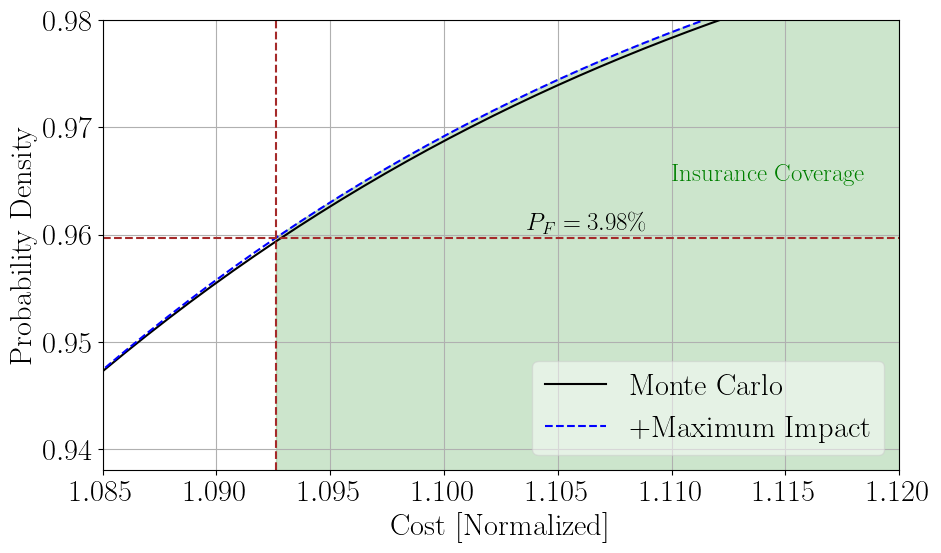}
    \caption{CDF of the optimal operational cost, with and without considering the maximum cost.}
    \label{fig:VaR}
\end{figure}

The PDF and CDF of the operational cost can be fitted using the results from the Monte Carlo simulation and the bi-level optimization. Fig.~\ref{fig:VaR} shows the CDF of the operational cost. At a confidence level $\alpha=3.98\%$, the horizontal coordinate of the intersection points represents the VaR under different fittings. It can be observed that the fitting considering the maximum impact results in a higher VaR, making the insurance policy more conservative. Furthermore, Table~\ref{Table:TVaR} presents the VaR and TVaR of the proposed grid model and insurance policy. If the grid operator purchases the insurance with a premium of \euro $0.29k$, the insurer will cover the operational costs exceeding \euro $68.06k$. 

Consider a scenario where the grid is highly vulnerable to attacks. Attackers can exploit the uncertainty of load variations, causing the grid to reach its maximum operational cost. In this scenario, according to the SMP framing in \ref{subsection:SMP}, $P_{GV}$ and $P_{VF}$ increase, resulting in a higher probability of failure $P_F$. We consider $P_F=3.98\%$ interpreted as the likelihood of an attack occurring, and the confidence level $\alpha$ is set to $5\%$~\cite{lau2021coalitional}. The results, shown in Table~\ref{Table:TVaR2}, indicate a significant increase in TVaR, with also a noticeable rightward shift in the starting point of the insurance coverage range, i.e., the VaR threshold. This is due to the higher probability of malicious cyber-attacks (LAAs) and a rise in operational costs compared to the previous fitting. This demonstrates how different risk modeling approaches can result in varied insurance policies and premiums. If the grid operator faces a higher risk of attacks due to insufficient defense mechanisms, the cyber insurance solution may become prohibitively expensive. Consequently, grid operators are more strongly encouraged to implement advanced defense mechanisms. On the other hand, if the system is equipped with advanced defense mechanisms, the insurance premium can be reduced.

From a cost-benefit perspective, the proposed framework can assist grid operators in evaluating the profitability of enhancing existing defense mechanisms compared to purchasing the proposed cyber-insurance coverage. In particular, the cost of upgrading defenses, in the worst-case scenario, must be equal to the proposed insurance premium; otherwise, it is more beneficial for the operator to purchase the insurance instead.

\begin{table}[!t]
    \centering
    \caption{Comparison of VaR, TVaR, and Premium under Different Risk Scenarios}
    \begin{minipage}[t]{0.48\textwidth}
        \centering
        \text{(a) Standard Risk Scenario}
        \begin{tabular}{ccc}
            \toprule
             & Percentage (\%) & Price ($k$\euro) \\
             & Nominal$=100\%$ & Nominal$=62.33k$\euro \\
            \midrule
            VaR     & 109.30 & 68.06 \\
            TVaR    & 11.85  & 1.22  \\
            Premium & 0.471  & 0.29  \\
            \bottomrule
        \end{tabular}
        \label{Table:TVaR}
    \end{minipage}
    \begin{minipage}[t]{0.48\textwidth}
        \centering
        \vspace{1mm}
        \text{(b) Highly Vulnerable to LAA ($P_{LAA}=3.98\%$)}
        \begin{tabular}{ccc}
            \toprule
             & Percentage (\%) & Price ($k$\euro) \\
             & Nominal$=100\%$ & Nominal$=62.33k$\euro \\
            \midrule
            VaR     & 112.70 & 71.38 \\
            TVaR    & 17.12  & 10.72 \\
            Premium & 0.856  & 0.53  \\
            \bottomrule
        \end{tabular}
        \label{Table:TVaR2}
    \end{minipage}
\end{table}

\subsection{Scalability Verification on IEEE European LV Network}

\begin{figure}[!t]
    \centering
    \subfigure[]{
        \includegraphics[width=0.4\textwidth]{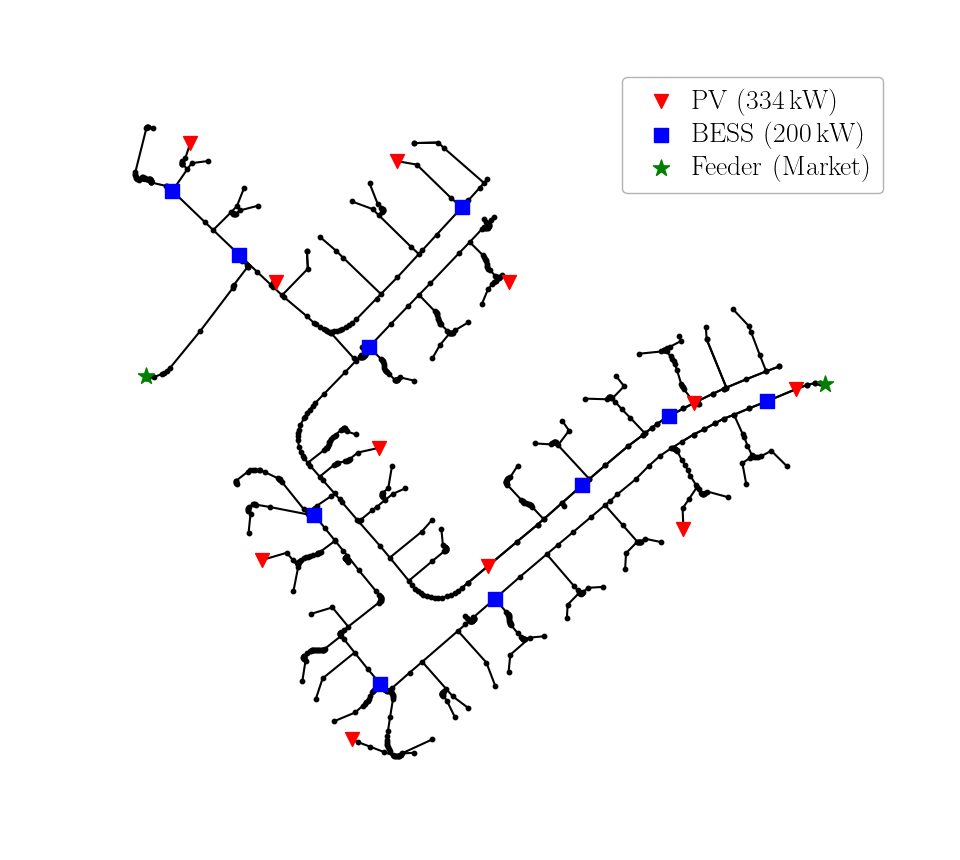}
        \label{fig:Topology906}
    }\\
    \subfigure[]{
        \includegraphics[width=0.225\textwidth]{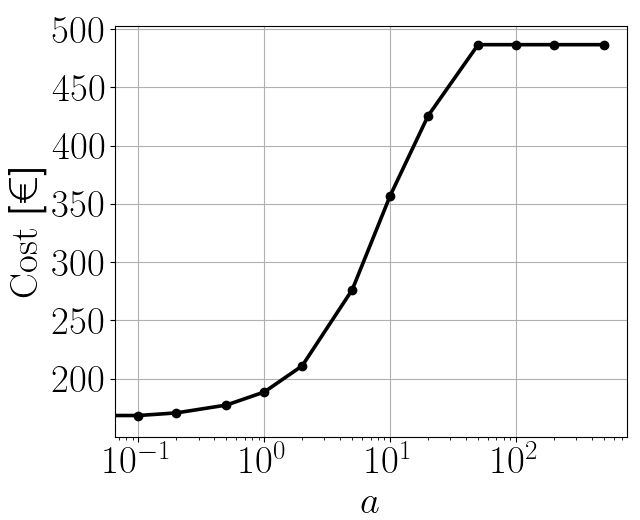}
        \label{fig:CostVSalpha906}
    }
    \hfill
    \subfigure[]{
        \includegraphics[width=0.225\textwidth]{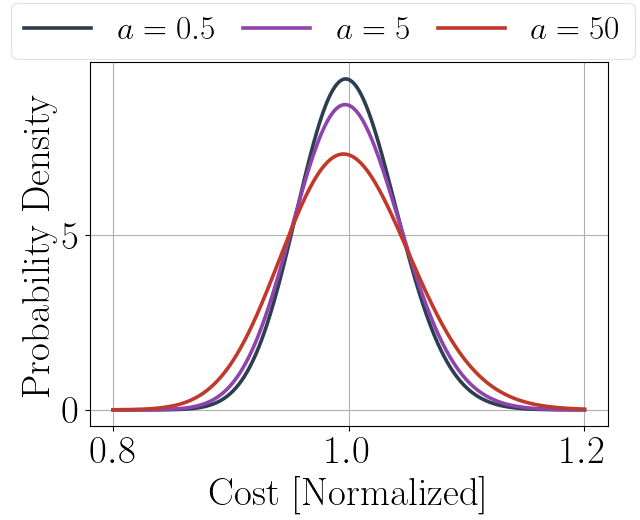}
        \label{fig:MonteCarloAlpha906}
    }
    \subfigure[]{
        \includegraphics[width=0.225\textwidth]{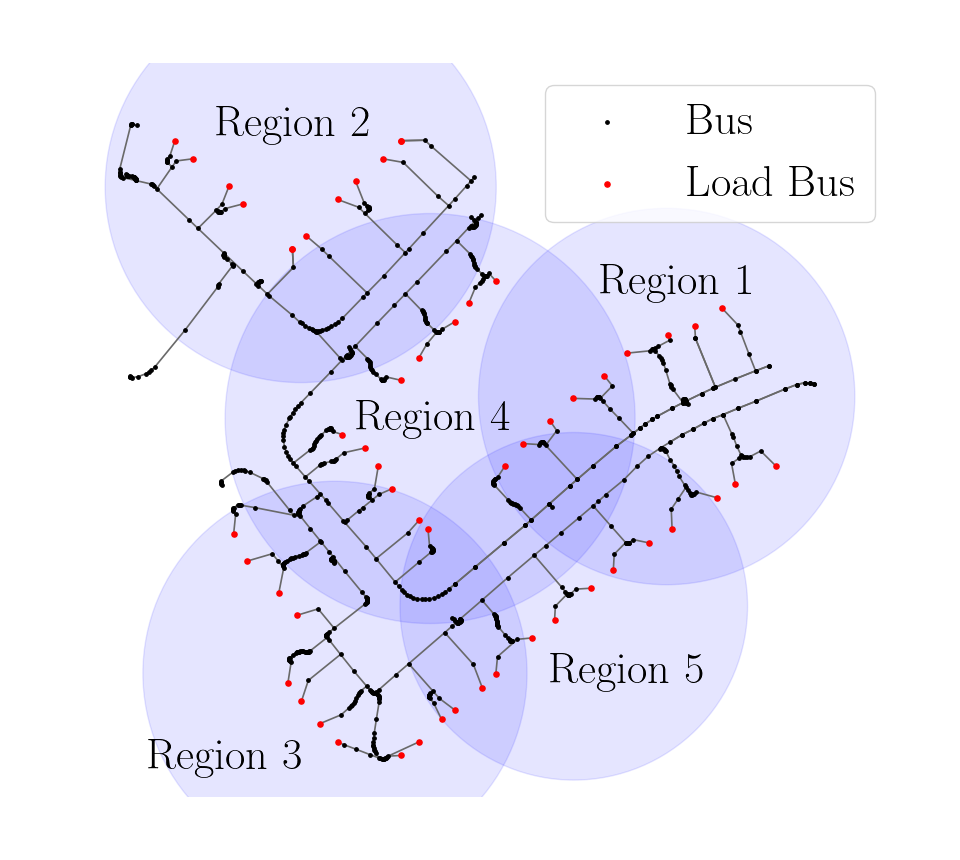}
        \label{fig:LAARegion906}
    }
    \hfill
    \subfigure[]{
        \includegraphics[width=0.225\textwidth]{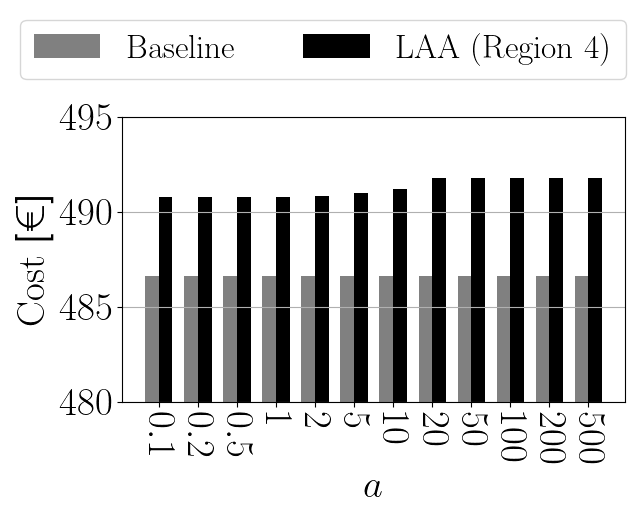}
        \label{fig:CostVSLAA906}
    }
    \caption{Scalability analysis of the proposed framework on the IEEE European LV test system: (a) Topology of the modified IEEE 906-bus system with PV, BESS, and electricity markets. (b) Total operational cost under different $a$ using cost function $\boldsymbol{c}^{LC}$ with $b=0$. (c) PDF of optimal operational cost with different $a$, $\delta=30\%$, $\Delta=10\%$. (d) Considered LAA regions of the IEEE 906-bus system. (e) Cost impact of an LAA at 8 p.m. in Region 4 under different $a$. }
    \label{fig:Scalability906}
\end{figure}

The second case study considers the IEEE European low-voltage test feeder to evaluate the scalability and practical feasibility of the proposed cyber insurance policy within a more realistic operational context. Fig.~\ref{fig:Topology906} depicts the network topology. The testbed system has altogether 906 buses and 55 loads. 
In the case study, we consider $12$ buses with $334$kW capacity of PVs and $10$ buses with a $200$kW BESS.
The hourly load profile is obtained by aggregating all minute-level load measurements whose timestamps fall within the specific hour, with a resulting average total system load of around $1.37$ MW per hour, peaks at $5.03$ MW.
The other system parameters are the same as the setup in \cite{IEEE_European_LV_Test_Feeder_v2}.

Fig.~\ref{fig:CostVSalpha906} illustrates the total operational cost under the load curtailment compensation function $\boldsymbol{c}^{LC}$ for different values of the penalty coefficient $a$. Similar to the IEEE 118-bus system, the operational cost increases significantly with the increase of $a$. The cost stops increasing when $a = 50$, because there are no voltage violations in our case study of the 906-bus system. Under this condition, further increases in the penalty coefficient $a$ do not trigger additional corrective actions, i.e., load curtailment, since the system remains well within its operational constraints. As a result, the total operational cost reaches a plateau, indicating that the economic response to more aggressive load curtailment penalties becomes saturated beyond this point.

To evaluate the distributional characteristics of the system optimal operational cost under uncertainty, Fig.~\ref{fig:MonteCarloAlpha906} presents the PDF obtained through Monte Carlo simulations with different values of $a$. When $a$ is small, the distribution is more concentrated, indicating that the system can effectively deal with load variations with minimal cost change due to its higher load curtailment flexibility. As $a$ increases, the system becomes less tolerant to load variations, resulting in broader and heavier-tailed cost distributions. This increased spread reflects a higher probability of extreme operational costs. Consequently, both $VaR_\alpha$ and the $TVaR_\alpha$ rise, since the quantitative and tail-average cost levels shift upward. 

We conduct LAA simulations across all predefined regions shown in Fig.~\ref{fig:LAARegion906} to evaluate the system's vulnerability to coordinated load manipulations. Among all the regions, Region 4 exhibits the most pronounced impact on operational cost, and is illustrated in Fig.~\ref{fig:CostVSLAA906}. The results demonstrate that LAAs can elevate system operational costs. The extent of this impact depends on the penalty coefficient $a$ in the load curtailment compensation function. When $a$ is small, the system allows more flexibility through load curtailment, which helps absorb the disruptive effects of the attack. As $a$ increases, the cost of curtailment becomes prohibitive, reducing the system’s ability to respond adaptively and thereby amplifying the economic consequences of the attack. When this cost amplification is reflected in the risk distribution, it leads to higher $VaR_\alpha$ and $TVaR_\alpha$ values, which in turn increase the insurance premium.

\section{Conclusion}
\label{section:Conclusion}

This paper proposes a cyber insurance policy that shifts part of the financial risk arising from extreme load variations, from grid operators to an insurer, thereby safeguarding the insured party from elevated operation costs. The proposed insurance framework considers both load variability and the system failure probability to determine the insurance coverage and premium. Beyond grid operation constraints, it also incorporates load curtailment penalties, resulting in a more realistic and practical insurance offering. We highlight that the proposed cyber insurance policy complements existing defense mechanisms and provides an additional protection layer to grid operators against undetected cyber-attacks in the form of LAAs. Notably, effective defense mechanisms reduce the system failure probability, thereby leading to a lower insurance premium. The effectiveness of the proposed insurance policy is evaluated through two case studies. Remarkably, with a premium of $0.47\%$ of the system's operational cost, the grid operator can hedge against additional operational costs, exceeding $9.3\%$, attributed to malicious load manipulations. The results demonstrate how the proposed cyber insurance policy can reduce financial risks and strengthen grid operators' ability to handle severe load variations and LAAs.

\section*{Acknowledgment}
The authors acknowledge the financial support of King Abdullah University of Science and Technology (KAUST).

\bibliographystyle{IEEEtran}
\bibliography{refs}
\end{document}